\documentclass[12pt]{article}
\usepackage{amsmath}
\usepackage{graphicx}
\usepackage{enumerate}
\usepackage{natbib}
\usepackage{url} 

\usepackage{ifthen}

\usepackage{mathrsfs}
\usepackage{amsmath,amssymb}
\usepackage{bm}
\usepackage{natbib}
\usepackage[usenames]{color}
\usepackage{amsthm}
\usepackage{subfigure}
\usepackage{multirow} 
\usepackage{enumitem}
\usepackage{caption}
\usepackage{subcaption}
\usepackage{enumitem}
\usepackage{dsfont}
\usepackage{mathtools}
\usepackage{caption}
\usepackage{booktabs}
\usepackage{array}
\usepackage[colorlinks,
linkcolor=red,
anchorcolor=blue,
citecolor=blue
]{hyperref}

\usepackage{mylatexstyle}
\ifdefined\final
\usepackage[disable]{todonotes}
\else
\usepackage[textsize=tiny]{todonotes}
\fi
\newcommand{\yl}[1]{\textcolor{black}{#1}}

\newcommand{\blind}{0}

\newif\ifincludeappendix
\includeappendixfalse

\begin{document}

\def\spacingset#1{\renewcommand{\baselinestretch}%
{#1}\small\normalsize} \spacingset{1}


\if0\blind
{
  \title{\bf Statistical Inference on High Dimensional Gaussian Graphical Regression Models}
  \author
{
Xuran Meng\thanks{Department of Biostatistics, University of Michigan; e-mail: {\tt xuranm@umich.edu}}
	~~and~~
Jingfei Zhang\thanks{Department of  Information Systems and Operations Management, Emory University;
  e-mail: {\tt emma.zhang@emory.edu}}
~~and~~
Yi Li\thanks{Department of Biostatistics, University of Michigan;
  e-mail: {\tt yili@umich.edu}}
}
  \maketitle
} \fi

\if1\blind
{
  \bigskip
  \bigskip
  \bigskip
  \begin{center}
    {\LARGE\bf Statistical Inference on High Dimensional Gaussian Graphical Regression Models}
\end{center}
  \medskip
} \fi

\bigskip
\begin{abstract}
Gaussian graphical regressions have emerged as a powerful approach for regressing the precision matrix of a Gaussian graphical model on covariates, which, unlike traditional Gaussian graphical models, can help determine how graphs are modulated by high dimensional subject-level covariates, and recover both the population-level and subject-level graphs. To fit the model, a multi-task learning approach {achieves} 
lower error rates compared to node-wise regressions. However, due to the high complexity and dimensionality of the Gaussian graphical regression problem, the important task of statistical inference remains unexplored. We propose a class of debiased estimators based on multi-task learners for statistical inference in Gaussian graphical regressions. We show that debiasing can be performed quickly and separately for the multi-task learners. 
In a key debiasing step {that estimates} 
the inverse covariance matrix, we propose a novel {projection technique} 
that dramatically reduces computational costs {in optimization} to scale only with the sample size $n$. We show that our debiased estimators enjoy a fast convergence rate and asymptotically follow a normal distribution, enabling valid statistical inference such as constructing confidence intervals and performing hypothesis testing. Simulation studies confirm the practical utility of the proposed approach, and we further apply it to analyze gene co-expression graph data from a brain cancer study, revealing meaningful biological relationships. 
\end{abstract}

\noindent%
{\it Keywords:} Gaussian graphical models, graphical model with covariates, multi-task learning, debiased inference, projection. 

\vfill


\newpage
\spacingset{1.9} 
\section{Introduction}
\label{sec:intro}
Gaussian graphical models  are powerful tools for describing dependencies among response variables in a homogeneous population \citep{meinshausen2006high, peng2009partial}. In precision medicine where subject-specific {gene} co-expression graphs are of interest, scenarios where the precision matrix depends on external covariates are  considered \citep{saegusa2016joint}.
For instance,  genetic variants and environmental factors affect both individual genes and their co-expression relationships in gene co-expression graphs \citep{wang2012snpxge2}; external factors like gender, age and genetic variants influence functional connectivity between brain regions \citep{zhang2023generalized}.
These applications have led to the development of covariate-dependent Gaussian graphical regressions, where the goal is to understand how external covariates modulate the graphical structures at the individual level and  recover both the population- and individual-level graphs. 
Towards this goal, \citet{ni2019bayesian} proposed a conditional graphical regression model that allows the structure of a directed graph to flexibly vary with continuous covariates under the assumption of a small number of covariates and a known hierarchical ordering of the nodes. \citet{zhang2023high} proposed a flexible Gaussian graphical regression framework that models the precision matrix as a function of covariates in high-dimensional settings, providing consistent estimators using sparse group lasso and separate node-wise regressions. Building on this, \citet{doi:10.1080/10618600.2024.2421246} further improved {the estimation procedure by proposing a joint multi-task learning problem, solved by an efficient augmented Lagrangian algorithm.}

Despite the advantages of Gaussian graphical regression approaches in recovering both population- and individual-level graphs, the existing work focused primarily on estimation, leaving a significant gap in statistical inference, especially in high dimensional settings. Statistical inference {tools are crucial, as they play} a fundamental role in quantifying the uncertainty of the estimated relationships by providing confidence intervals and enabling hypothesis testing. In this paper, we address this gap by developing computationally efficient methods for conducting statistical inference in Gaussian graphical regressions.

In high-dimensional problems, a useful approach to statistical inference involves debiasing methodologies, as it is well-known that estimators like the lasso 
suffer from {non-negligible} 
bias in their second-order expansions, leading to inaccurate results in inference \citep{tibshirani2005sparsity, zou2006adaptive, bellec2022biasing, cai2022sparse}. 
Using debiasing algorithms in the lasso framework, \citet{javanmard2014confidence} and \citet{van2014asymptotically} developed confidence intervals and $p$-values for high-dimensional linear regressions. 
\citet{chernozhukov2018double} introduced Double Machine Learning (DML), which employs Neyman-orthogonal moments and cross-fitting to handle high-dimensional nuisance parameters in semiparametric models. \citet{zhu2018linear} explored debiasing in linear models without assuming sparsity. \citet{fei2019drawing} and \citet{fei2021estimation} enhanced inference methods through sample-splitting for improved precision. A series of works \citep{zhang2019optimal, zhang2020islet, xia2022inference} also tackled the computational challenges in tensor regression, developing unbiased algorithms tailored to low-rank tensor structures. 

Estimation and statistical inference in high-dimensional Gaussian graphical regressions present several {new} challenges, including {the high model complexity, dimensionality and significant computational burden.} 
{From a computational perspective, a multi-task learning estimator that simultaneously solves for all $O(p^2q)$ coefficients, where $p$ represents the number of Gaussian variables and $q$ the number of covariates, enjoys a fast convergence rate. 
Although parameter estimation can be efficiently handled using an augmented Lagrangian algorithm \citep{doi:10.1080/10618600.2024.2421246}, debiasing all $O(p^2q)$ coefficients simultaneously remains  computationally demanding.}
{From a theoretical perspective, the design matrix in the Gaussian regression problems includes interaction terms}
between the {$p$} Gaussian variables and the {$q$} external covariates, where the Gaussian variables are  also dependent among themselves.
Consequently, each row of the design matrix has a very complex joint distribution and is not sub-Gaussian or sub-exponential, as is commonly assumed in the existing literature \citep{cai2022sparse}. Moreover, unlike standard multi-task learning problems where the stochastic terms from separate regressions  are independent, 
the $p$ regression tasks in the graphical setting inherently exhibit  a complex dependence structure influenced by external covariates, making {a} joint inference virtually infeasible. 


We address these challenges as follows. First, to overcome the theoretical difficulties of joint inference in standard multi-task learning, {our key finding is that although the coefficients are estimated jointly, they can be debiased separately. Specifically,} we propose a marginalized debiasing procedure that decomposes the inference of the full vector into segments, ensuring that statistical validity is maintained within each segment (i.e., within each node). Second, unlike traditional debiasing methods \citep{javanmard2014confidence, cai2022sparse}, which exhibit polynomial growth in {optimization} costs with $p$ and $q$, we introduce a new {projection} 
technique that ensures computational costs in optimization scale with the sample size $n$. {Specifically, our approach maps the constrained optimization problem from $\mathbb{R}^{(p-1)(q+1)}$ to $\mathbb{R}^{n}$, quickly solves the much smaller problem, and then maps the solution back to the larger space, leading to a significant reduction in computational cost.} To our knowledge, this is the first approach to enable reliable statistical inference in high-dimensional Gaussian graphical regression models with combined penalty functions. We provide theorems for asymptotic inference and validate our approach through simulations. Moreover, our results also notably relax a condition on the number of nonzero coefficients in existing work \citep{doi:10.1080/10618600.2024.2421246,zhang2023high} from $O(n^{1/6})$ to $o(n^{1/2})$.

More closely related to our work, \citet{hudson2022covariate} explored a debiasing method for covariate-adjusted testing in differential graph analysis, {but their approach is limited to a small number of variables;} \citet{cai2022sparse} considered debiased sparse group lasso, but their method is not directly applicable in our settings as our design matrix includes high-dimensional interaction terms. {Furthermore, the computational cost of their method can be prohibitive under our setting as it increases polynomially with $p$ and $q$.}

The paper is organized as follows. In Section~\ref{sec:pre}, we describe the Gaussian graphical regression model and review its multi-task estimators. Section~\ref{sec:main_result} proposes a new inference method for Gaussian graphical regression models, by deriving a debiased estimator via projection technique and establishing the theoretical results for statistical inference.  Section~\ref{sec:simulations} conducts simulations to examine the finite sample performance of the method. Section~\ref{sec:real_data} applies our inference approach to analyze a brain tumor data set to study how SNPs may influence gene co-expression graphs and reports biologically meaningful results.  Section~\ref{sec:conclusions} concludes the paper. Additional simulations and proofs are provided in the supplement.

\section{The Preamble}
\label{sec:pre}
 
We first present notation. Throughout the paper, lowercase letters denote scalars, and boldface denotes vectors and matrices.
The sets of natural and real  numbers are denoted by $\NN$ and  $\RR$, respectively. 
We denote $[n_1:n_2]=\{n_1,\ldots,n_2\}$ and 
$[n]=[1:n] =\{1,\ldots,n\}$.  We use ``$\odot$'' to denote the Hadamard product. 
We write $X_1(n)=O(X_2(n))$ or  $X_1(n)\precsim X_2(n)$ if for any
$\varepsilon>0$, there exists $C>0$ such that $\PP(|X_1(n)/X_2(n)|>C)\leq \varepsilon$ for all $n$. We denote $X_1(n)=o(X_2(n))$ if $\{X_1(n)/X_2(n)\}$ converges to $0$ in probability. We use $\|\cdot\|_q$ to denote the $\ell_q$ Euclidean norm.
Let $\bgamma=(\bgamma^\top_{(1)},\ldots,\bgamma^\top_{(q)})$ be a vector associated with a pre-defined group structure, where $\bgamma_{(j)}\in\mathbb{R}^{p_j}$ is the sub-vector corresponding to group $j$ for $j\in[q]$. Then, the $\ell_{q_1,q_2}$-norm of $\bgamma$ is defined as: $\|\bgamma\|_{q_1,q_2} = \left( \sum_{j} \|\bgamma_{(j)}\|_{q_2}^{q_1} \right)^{1/q_1}$, 
for $0 \leq q_1, q_2 \leq+\infty$.
For example, $\|\bgamma\|_{\infty,2} = \max_{j} \|\bgamma_{(j)}\|_2$.

\subsection{Gaussian graphical regression} 
\label{subsec:intro_GGM}
We {briefly review} 
Gaussian graphical models by considering a random vector $\bX = (X_1, \ldots, X_p)$, which follows a multivariate normal distribution $\cN_p(\zero, \bSigma)$ and $\bSigma \in \RR^{p \times p}$ is the covariance matrix. The inverse of $\bSigma$, denoted as $\bOmega=\bSigma^{-1} = (\sigma^{ij})_{p \times p}$, is the precision matrix, where  $\sigma^{ij} \neq 0$ implies that $X_i$ and $X_j$ are conditionally dependent given the other variables. 
To estimate these conditional dependencies, \citet{meinshausen2006high} and \citet{peng2009partial} established the dependence 
of each variable $X_j$ on the remaining variables $\bX_{-j} = (X_1, \dots, X_{j-1}, X_{j+1}, \dots, X_p)$ via
\begin{align}
X_j = \sum_{k \neq j}^p \beta_{jk} X_k + \varepsilon_j,\label{eq:basic_regression}
\end{align}
where $\varepsilon_j$ is independent of $\bX_{-j}$ if and only if $\beta_{jk} = -\sigma^{jk}/\sigma^{jj}$. Estimating the nonzero $\beta_{jk}$ coefficients thus becomes equivalent to identifying nonzero $\sigma^{ij}$ elements. 

Recently, \cite{zhang2023high,doi:10.1080/10618600.2024.2421246} considered covariate-dependent Gaussian graphical models, {where $\bU = (U_1, \dots, U_q)^\top$ represents a $q$-dimensional vector of covariates and the} conditional distribution of $\bX$ given $\bU = \ub$ is modeled as:
\begin{align}  \label{eq:multi-gauss}
    \bX \mid \bU = \ub \sim \cN_p(\bmu(\ub), {\bOmega^{-1}}(\ub)).
\end{align}
The mean vector $\bmu(\ub)$ and {precision matrix $\bOmega(\ub)$} 
are given by 
\begin{align}
\bmu(\ub)=\bGamma\ub,\quad 
\bOmega(\ub) = \Bb_0 + \sum_{h=1}^q \Bb_h u_h,\label{eq:Omega_form}
\end{align}
where $\bGamma\in\RR^{p\times q}$, and  $\Bb_0, \Bb_1, \dots, \Bb_q \in \RR^{p \times p}$ are symmetric matrices.
In this framework, the {covariate effect} 
is expressed through the contribution of each $\Bb_h$ to the precision matrix. 
{The diagonal elements of $\bOmega(\ub)$ are usually assumed to be $\Omega(\ub)_{jj} = \sigma^{jj}$ so that the residual variance of $X_j$ does not vary with covariates, making the analysis more tractable.}
The formulation of \eqref{eq:multi-gauss} and 
\eqref{eq:Omega_form} 
is useful as, after centering the vector $\bZ = \bX - \bGamma\ub = (Z_1, \dots, Z_p)^\top$, they can be rewritten as:
\begin{align}
Z_j = \sum_{k \neq j}^p \beta_{jk0} Z_k + \sum_{k \neq j}^p \sum_{h=1}^q \underbrace{\beta_{jkh} \cdot u_h \cdot Z_k}_{\text{Interaction term}} + \varepsilon_j, \label{eq:basic_regression2}
\end{align}
where $\beta_{jkh} = -(\Bb_h)_{jk} / \sigma^{jj}$, $\epsilon_j$ is independent of $\bZ_{-j}$ and $\text{Var}(\epsilon_j)=1/\sigma^{jj}$, for all $j,k$ and $h$. As \eqref{eq:basic_regression2} offers a regression framework for estimating the dependence of precision parameters in \eqref{eq:Omega_form} on  $\ub$,  it is termed Gaussian graphical regression on external covariates. It generalizes \eqref{eq:basic_regression} by incorporating interactions between $\bZ_{-j}$ and $\ub$, and, correspondingly, modeling the partial correlation between $Z_j$ and $Z_k$  as a function of $\ub$.

\subsection{Multi-task learning for Gaussian graphical regressions} 
\label{sec:multi/multi_learning}
Consider $n$ independent observations, $(\ub^{(i)},\xb^{(i)})\in\RR^{q}\times \RR^{p}$, where $i\in[n]$. Let $\zb^{(i)}=\xb^{(i)}-\bGamma\ub^{(i)}$. 
{To expose the key ideas, we assume $\bGamma$ is known in the ensuing development and focus on the estimation and inference of $\beta_{jhk}$'s.}
Letting $\zb_j=(z_j^{(1)},...,z_j^{(n)})^\top$ for $j\in[p]$ and $\ub_h=(u_h^{(1)},...,u_h^{(n)})^\top $ for $h\in[q]$, the Gaussian graphical regression model on the $j$th response variable can be written as 
\begin{align*}
    \zb_j=\sum_{k \neq j}^p \beta_{j k 0} \zb_k+\sum_{k \neq j}^p \sum_{h=1}^q \beta_{j k h} \ub_h \odot \zb_k+\bvarepsilon_j.
\end{align*}
Here, $\bvarepsilon_j\in\RR^n$ is the regression error vectors of the node $j$ with $\bvarepsilon_j\sim \cN(0,\frac{1}{\sigma^{jj}}\Ib)$. 
We write $\bbeta_j=((\bbeta_j)_{(0)}^\top, ...,(\bbeta_j)_{(q)}^\top)^\top \in \RR^{(p-1)(q+1)}$, where $(\bbeta_j)_{(h)}=(\beta_{j1h}, \dots, \beta_{jph})^\top \in \RR^{p-1}$ groups all the coefficients in the $h$-th group. A more detailed form of $\bbeta_j$ is as follows:
\begin{align*}
\bbeta_j=(\underbrace{\beta_{j10}, \dots, \beta_{jp0}}_{(\bbeta_j)_{(0)}}, \quad \underbrace{\beta_{j11}, \dots, \beta_{jp1}}_{(\bbeta_j)_{(1)}}, \quad \dots, \quad \underbrace{\beta_{j1q}, \dots, \beta_{jpq}}_{(\bbeta_j)_{(q)}})^\top.
\end{align*}
Let $\bbeta = (\bbeta_1^\top, \dots, \bbeta_p^\top)^\top\in\RR^{p(p-1)(q+1)}$, and collect all group vectors $(\bbeta_j)_{(h)}$ defined above into the vector $\bbb_h$, where $\bbb_h = ((\bbeta_1)_{(h)}^\top, \dots, (\bbeta_p)_{(h)}^\top)^\top\in\RR^{p(p-1)}$. To ease notation, we define a large design matrix $\cW \in \RR^{np \times p(p-1)(q+1)}$ as: 
\begin{align*}
\cW = \begin{pmatrix}
\Wb_1 & \cdots & \zero_{n \times (p-1)(q+1)} \\
\vdots & \ddots & \vdots \\
\zero_{n \times (p-1)(q+1)} & \cdots & \Wb_p
\end{pmatrix},
\end{align*}
where $\Wb_j=[\zb_1,...,\zb_{j-1},\zb_{j+1},...,\zb_p,\zb_1\odot\ub_1,...,\zb_{j-1}\odot\ub_1,\zb_{j+1}\odot\ub_1,...,\zb_p\odot\ub_q]\in\RR^{n\times (p-1)(q+1)}$, and the response vector $\yb$ as $\yb=(z_1^{(1)},...,z_1^{(n)},z_2^{(1)},...,z_p^{(n)})^\top\in\RR^{np}$. 
The \textbf{multi-task learning} {simultaneously estimates all $\bbeta_j$ by minimizing the following loss function:}
\begin{align}
    \hbbeta=\argmin_{\bbeta} \frac{1}{2n}\big\|\yb-\cW\bbeta\big\|_2^2+\lambda_e\|\bbeta\|_1+\lambda_g\sum_{h=1}^q \|\bb_h\|_2,\label{eq:multi_estimator}
\end{align}
where $\lambda_e,\lambda_g>0$ are two tuning parameters. 
The regularization in \eqref{eq:multi_estimator} is known as the \textit{sparse group lasso penalty} \citep{simon2013sparse,li2015multivariate,cai2022sparse,zhang2023high} because it combines both element- and group-level sparsity constraints; the term, $\|\bbeta\|_1$, promotes element-wise sparsity, since effective covariates may affect only a few edges; the term $\sum_{h=1}^q \|\bb_h\|_2$ encourages group-wise sparsity, with the exception of $h=0$, where group sparsity is not enforced. 
{As each group coefficient vector $\bb_h$ collects $(\bbeta_1)_{(h)},\ldots,(\bbeta_p)_{(h)}$ from all $p$ tasks, }\eqref{eq:multi_estimator} defines a multi-task learning framework by considering $p$ graphical regressions simultaneously as well as a group lasso penalty to {regularize} 
$\bb_h$ across regressions. This enables us to borrow information across the $p$ tasks to select and estimate effective covariates. 
{Indeed, \cite{doi:10.1080/10618600.2024.2421246} showed that the error rate of $\hbbeta$ may improve over estimators from separate regressions by a factor of $p$.} The estimation problem in \eqref{eq:multi_estimator} can be solved efficiently via a computationally efficient Fenchel convexification \citep{mifflin1977semismooth,doi:10.1080/10618600.2024.2421246}.

\section{Segmentally Debiased Multi-task Graphical Regression via Projection}
\label{sec:main_result}
As $\hbbeta$ underestimates the coefficients due to the shrinkage effects of the penalty function, it is critical to debias $\hbbeta$ for valid statistical inference. 
{While the multi-task learning estimator in \eqref{eq:multi_estimator} achieves a fast convergence rate, debiasing the entire vector $\hbbeta$ simultaneously presents significant theoretical challenges. Unlike in traditional multi-task learning problems, the $p$ regression tasks in \eqref{eq:multi_estimator} are not independent and exhibit a highly complex dependence structure. Specifically, $\Cov(\varepsilon_j,\varepsilon_{j'})$ depends on $\ub$, and the exact form of this relationship is intractable. Hence, }
directly calculating $\hbbeta$'s joint distribution is nearly impossible. 
{Our key finding to addressing this problem is that, although the coefficients are estimated jointly, they can be debiased separately. Specifically,}
we propose to decompose the inference of the entire $\bbeta$ vector into segmental inferences. This segmental debiasing method further reduces computational complexity and enables valid statistical inference within each segment, avoiding the need to account for dependencies across tasks. This approach is conceptually similar to the marginal method, as seen in settings like generalized estimating equations (GEE) \citep{ziegler2010generalized}.

Specifically, we segment the debiasing procedure of $\hbbeta$ to each component $\hbbeta_j$ individually, 
i.e., for each $j \in [p]$, we debias $\hbbeta_j$ separately and denote the  estimator by $\hbbeta_j^{u}$:
\begin{equation}
\hbbeta_j^{u} = \hbbeta_j + \frac{1}{n} \hat{\Mb}_j^\top \Wb_j^\top (\zb_j - \Wb_j \hbbeta_j).\label{eq:multi_learning_debiase_estimator}
\end{equation}
We  refer to \eqref{eq:multi_learning_debiase_estimator} as the {\em \underline{s}egmentally \underline{a}djusted  \underline{g}raphical r\underline{e}gression} (SAGE)
estimator. We 
{later show the SAGE estimator $\hbbeta_j^{u}$ asymptotically follows a multivariate normal distribution, laying a foundation for inference. 
In \eqref{eq:multi_learning_debiase_estimator}, $\hat{\Mb}_j = [\hat{\mb}_{j1}, \dots, \hat{\mb}_{j(p-1)(q+1)}]$ is an estimator of $\bSigma_{\Wb_j}^{-1}$, where $\bSigma_{\Wb_j} = \EE~\Wb_j^\top \Wb_j / n$. Let $\hbSigma_{\Wb_j} = \Wb_j^\top \Wb_j / n$. In existing methods \citep{javanmard2014confidence,cai2022sparse}, $\hat{\mb}_{jl}$'s are typically estimated by solving:
\begin{equation}
\label{eq:debias_estimate_inverse_m}
    \begin{split}
        &\argmin_{\mb\in\RR^{(p-1)(q+1)}} \mb^\top\hbSigma_{\Wb_j}\mb\\
        &\text{subject to}\quad\|H_{\alpha}(\hbSigma_{\Wb_j}\mb-\eb_l)\|_{\infty,2}\leq \gamma,
    \end{split}
\end{equation}
where the soft-thresholding operator $H_{\alpha}(x)=\sign(x)\cdot(|x|-\alpha)_+$ applies pointwise to vectors,  $\eb_l \in  \mathbb{R}^{(p-1)(q+1)}$ is the standard basis vector,   and the scalars $\alpha$ and $\gamma$ are to be defined via theoretical analysis. 
In \eqref{eq:debias_estimate_inverse_m}, the computing cost for estimating each $\hat{\mb}_{jl}$ in the space of $\mathbb{R}^{(p-1)(q+1)}$ increases polynomially with $p$ and $q$.

Alternatively, since $\hbSigma_{\Wb_j}$ has rank  $n$, this motivates us to map the optimization problem from $\mathbb{R}^{(p-1)(q+1)}$ to  $\mathbb{R}^{n}$, making computation feasible. Specifically, consider the $n \times n$ matrix $\bXi_j = \Wb_j \Wb_j^\top / n$, with its eigen-decomposition given by $\bXi_j = \Ub_j \Db_j \Ub_j^\top$, where $\Ub_j^\top\Ub_j=\Ub_j\Ub_j^\top=\Ib_n$ \citep{golub1971singular}. The diagonal matrix $\Db_j\in\mathbb{R}^{n\times n}$ collects the  eigenvalues of $\bXi_j$, which are also identical to the nonzero eigenvalues of $\hbSigma_{\Wb_j}$. Now, define $\Vb_j = \Wb_j^\top \Ub_j \Db_j^{-1/2} / \sqrt{n}$. It is then easy to derive that $\Wb_j/\sqrt{n}=\Ub_j\Db_j^{1/2}\Vb_j^\top$ and $\Vb_j^\top \Vb_j=\Ib_n$. Correspondingly, we have $\hbSigma_{\Wb_j} = \Wb_j^\top\Wb_j/n =\Vb_j\Db_j\Vb_j^\top$.
This allows us to consider the following optimization problem in  $\mathbb{R}^{n}$:
\begin{equation}
\label{eq:debias_estimate_inverse}
    \begin{split}
        & \argmin_{\btheta\in\mathbb{R}^n}\btheta^\top\Db_j\btheta,\\
        &\text{subject to}\quad\|H_{\alpha}(\Vb_{j}\Db_j\btheta-\eb_l)\|_{\infty,2}\leq \gamma,
    \end{split}
\end{equation} 
where 
$\alpha$ and $\gamma$ are defined later in our main results. 
Denote by $\hat{\btheta}_{jl}$, which solves \eqref{eq:debias_estimate_inverse}.
{As the columns of the orthonormal matrix $\Vb_j$ are eigenvectors of $\hbSigma_{\Wb_j}$
spanning an $n$-dimensional subspace, the $\btheta$ in \eqref{eq:debias_estimate_inverse} is a projection of $\mb$ in \eqref{eq:debias_estimate_inverse_m} onto the columns of $\Vb_j$.} 
{The following proposition shows the properties of $\hat{\btheta}_{jl}$ and $\hat{\mb}_{jl}$, justifying the utility of this projection, and its proof is in Section~\ref{sec:reformulation_debiase_optimize} of the supplement.
\begin{proposition}
\label{prop:debias_trans}  If $\hat{\btheta}_{jl}$ is a solution of \eqref{eq:debias_estimate_inverse}, then
$\hat{\mb}_{jl} = \Vb_j \hat{\btheta}_{jl}$ is a solution of \eqref{eq:debias_estimate_inverse_m}.
Inversely, if $\hat{\mb}_{jl}$ is the solution of \eqref{eq:debias_estimate_inverse_m}, then $\hat{\btheta}_{jl}=\Vb_j^\top\hat{\mb}_{jl}$ is a solution of \eqref{eq:debias_estimate_inverse}.
\end{proposition}}
The advantage of \eqref{eq:debias_estimate_inverse} lies in its estimation in $\mathbb{R}^{n}$, a space whose dimension does not depend on $p$ or $q$ and is typically  much lower than $(p-1)(q+1)$,  avoiding estimating the inverse  matrix directly in the original $\mathbb{R}^{(p-1)(q+1)}$ and saving much computation. Once $\hat{\btheta}_{jl}$ is computed, we can map it back to  $\mathbb{R}^{(p-1)(q+1)}$ to obtain the estimate of each column of the inverse matrix as $\hat{\mb}_{jl} = \Vb_j \hat{\btheta}_{jl}$, with $j \in [p-1)(q+1)]$. 
{Moreover,  the diagonal structure of 
$\Db_j$ simplifies the computation 
in \eqref{eq:debias_estimate_inverse} by
avoiding full matrix operations.} 
Thus, it is clear that our debiased method remains computationally efficient even as the dimension of $\bSigma_{\Wb_j}$ grows.  When estimating $\hat{\mb}_{jl}$, 
the computational cost with \eqref{eq:debias_estimate_inverse} is $O(n)$, whereas the direct applications of optimization in \eqref{eq:debias_estimate_inverse_m} incur a cost of $O(p^2q^2)$; see Section~\ref{sec:reformulation_debiase_optimize} in the supplement.
Importantly, the asymptotic results of  $\hbbeta_j^{u}$ of  \eqref{eq:multi_learning_debiase_estimator}, with $\hat{\Mb}_j$ estimated 
based on $\hat{\btheta}_{jl}$'s, 
 can be established  under the following assumptions.
\begin{assumption}
\label{assump:1}
Suppose $\ub^{(i)}$ are i.i.d mean zero random vectors with a covariance matrix satisfying $\phi_0\geq\lambda_{\max}(\Cov(\ub^{(i)}))\geq\lambda_{\min}(\Cov(\ub^{(i)}))\geq 1/\phi_0$ for some constant $\phi_0>0$. Moreover, there exists a constant $M>0$ such that $|u_h^{(i)}|\leq M$ for all $i$ and $h$.
\end{assumption}
 
\begin{assumption}
\label{assump:2}
Suppose that $\phi_1\leq\lambda_{\min}(\Cov(\zb^{(i)}))\leq \lambda_{\max}(\Cov(\zb^{(i)}))\leq \phi_2$ for some constants $\phi_1,\phi_2>0$. 
\end{assumption}

\begin{assumption}
\label{assump:3}
The dimensions $p$, $q$, and the element-wise sparsity $s_e$ satisfy ${s_e}(\log(p) + \log(q)) = o(\sqrt{n}/\log(n))$. {Additionally, the maximum column $\ell_0$ norm of $\bOmega(\ub)$ is  bounded above by a positive constant  $c>0$.}
\end{assumption}


Assumption~\ref{assump:1} is  characterizes the joint distribution of each row in $\Wb_j$. This condition is not restrictive, as genetic variants are often encoded as ${0,1}$ or ${0,1,2}$ \citep{chen2016asymptotically}. Similar assumptions can be found in \citet{doi:10.1080/10618600.2024.2421246, zhang2023high}. 
Assumption~\ref{assump:2} imposes bounded eigenvalues on $\Cov(\zb^{(i)})$. With Assumption~\ref{assump:1}, this provides a well-defined characterization of the joint distribution for each row in $\Wb_j$.  Assumption~\ref{assump:2} is mild and commonly used in the literature \citep{chen2016asymptotically, cai2022sparse, zhang2023high}. 
Assumption~\ref{assump:3} is also mild, as we assume $\sqrt{n} = \omega(s_e (\log p + \log q))$. This condition is less restrictive compared to \citet{doi:10.1080/10618600.2024.2421246}, where $s_e $, $ \log q$ and $\log q$ are required to grow no faster than $n^{1/6}$. Our assumption allows $s_e(\log p + \log q)$ to grow  more slowly than $\sqrt{n}$, providing more flexibility.


\begin{theorem}
\label{mainthm:Known_Gamma_multi}
Suppose $\bbeta\in\RR^{p(p-1)(q+1)}$ is $(s_e,s_g)$-sparse, Assumptions~\ref{assump:1}-\ref{assump:3} hold and $s_{\lambda}\cdot\log(pq)=O(\sqrt{n}/\log n)$,
where $s_{\lambda}$ is the number of nonzero entries in a candidate model such that $s_e<s_{\lambda}\leq n$
.  Set 
\begin{align*} \lambda_{e}=C\sqrt{\frac{2s_{e}\log(ep)+s_{g}\log(eq/s_{g})}{ns_{e}}},\quad \lambda_{g}=\sqrt{\frac{s_{e}}{s_{g}}}\lambda_e,
\end{align*}
and let $\alpha=C\sqrt{\log(pq)/n}$, $\gamma=\sqrt{s_{e}/s_{g}}\cdot\alpha$, then with probability at least $ 1 - C_1\exp(-C_2\{s_{e}\log(ep) + s_{g}\log(eq/s_{g})\}/{s_{e}}) $ for some constants $C_1,C_2>0$, the SAGE estimator $\hbbeta_j^{u}$ in \eqref{eq:multi_learning_debiase_estimator} can be decomposed into $\sqrt{n}(\hbbeta_j^{u}-\bbeta_j)=\bDelta_j+w_j$, where
\begin{align*}
\|\bDelta_j\|_{\infty}\leq \frac{Cs_e(\log(p)+\log(q))}{\sqrt{n}},\quad w_j|\Wb_j\sim\cN(0,\frac{1}{\sigma^{jj}}\hMb_j^\top\hbSigma_{\Wb_j}\hMb_j),
\end{align*}
and $\hMb_j$ is obtained from \eqref{eq:debias_estimate_inverse}. 
Moreover,  
for any $ l \in [(p-1)(q+1)]$ it holds that 
\begin{align*}
    \frac{\sqrt{n}\big((\hbbeta_j^{u})_l-(\bbeta_j)_l\big)}{\sqrt{\hmb_{jl}^\top \hbSigma_{\Wb_j}\hmb_{jl}}}
    \stackrel{d}{\to} \cN(0,\frac{1}{\sigma^{jj}}).
\end{align*}
\end{theorem}
The theorem and its proof provide several theoretical advances.
First, we have established the existence of a solution to the convex optimization problem in \eqref{eq:debias_estimate_inverse}
with suitable conditions for $\alpha$ and $\gamma$,
despite the complex joint distribution of rows in the design matrix $\Wb_j$, which are neither sub-Gaussian nor sub-exponential. To ensure feasibility of the solution in \eqref{eq:debias_estimate_inverse} and control the error term $\bDelta_j$, we reconstructed probability bounds tailored to this structure.
Second, we derived a lower bound for the term $\hmb_{jl}^\top \bSigma_{\Wb_{j}} \hmb_{jl}$, ensuring it stays away from zero and thus stabilizing the estimator. The complexity of $\Wb_j$ demanded advanced techniques beyond traditional concentration bounds, specifically using tools like 
 Lemma~\ref{lemma:from_kuchi}.  
Third, as \eqref{eq:multi_estimator} lacks a closed-form solution, the proof requires careful treatment as shown in Section~\ref{sec:proofthm_known_multi} of the supplement, particularly given our  more relaxed sparsity conditions compared to \citet{doi:10.1080/10618600.2024.2421246};  see Lemmas~\ref{lemma:known_gamma_multi_delta} and \ref{lemma:known_gamma_multi_delta1}.


Practically, this theorem enables us to construct confidence intervals. 
If $\sigma^{jj}$ were known,  an asymptotic $(1-\alpha)$-confidence interval of   $(\bbeta_j)_l $ would be
\begin{align} \label{CI}
    \left[(\hbbeta_j^{u})_l-\frac{\Phi^{-1}(1-\alpha/2)}{\sigma^{jj}}\sqrt{\frac{\hmb_{jl}^\top \hbSigma_{\Wb_j}\hmb_{jl}}{n}}, ~ (\hbbeta_j^{u})_l+\frac{\Phi^{-1}(1-\alpha/2)}{\sigma^{jj}}\sqrt{\frac{\hmb_{jl}^\top \hbSigma_{\Wb_j}\hmb_{jl}}{n}} ~ \right].
\end{align}
  As proposed by \citet{zhang2023high}, a natural and consistent estimator of $\frac{1}{\sigma^{jj}}$ is \newline
$    \frac{1}{\hat{\sigma}^{jj}}=\frac{1}{n-\hat{s}_j}\|\zb_j-\Wb_j\hbbeta_j^{OLS}\|_2^2, $
where $\hbbeta_j^{OLS}$ is the OLS estimator constrained on the set   $\hat{\cS}_j$,  satisfying  $ \Big(\hbbeta_j^{OLS}\Big)_{\hat{\cS}_j}=\big((\Wb_j)_{\hat{\cS}_j}^\top(\Wb_j)_{\hat{\cS}_j}\big)^{-1}(\Wb_j)_{\hat{\cS}_j}^\top\zb_j$ and $\Big(\hbbeta_j^{OLS}\Big)_{\hat{\cS}_j^c}=\zero$. Hence, we can construct the confidence interval for $(\bbeta_j)_l$ by replacing $\sigma^{jj}$
with $\hat{\sigma}^{jj}$ in (\ref{CI}). Moreover, for tests of linear contrasts, such as $(\bbeta_j)_1 = (\bbeta_j)_2$, the following corollary addresses the inference for $\Ab\bbeta_j$ for  fixed matrix  $\Ab$. It forms the basis for testing, for example, whether a QTL modulate co-expressions in a gene pathway consisted of a set of genes.

\begin{corollary}
\label{coro:2}
Under the same condition of Theorem~\ref{mainthm:Known_Gamma_multi}, for any fixed matrix $\Ab\in\RR^{K\times (p-1)(q+1)}$ for some $K\in\NN$, it holds with probability at least $ 1 - C_1\exp(-C_2\{s_e\log(ep) + s_{g}\log(eq/s_{g})\}/{s_{e}})$ for some constant $C_1,C_2>0$, the  vector $\Ab\hbbeta_j^{u}-\Ab\bbeta_j$ can be decomposed into $\sqrt{n}(\Ab\hbbeta_j^{u}-\Ab\bbeta_j)=\bDelta_j+w_j$, where
\begin{align*}
    \|\bDelta_j\|_{\infty}\leq \frac{C\|\Ab\|_{\infty}s_e(\log(p)+\log(q))}{\sqrt{n}}, \quad w_j\mid\Wb_j\sim\cN_s(0,\frac{1}{\sigma^{jj}}\Ab\hMb_j^\top\hbSigma_{\Wb_j}\hMb_j\Ab^\top),
\end{align*}
with $\hMb_j$  obtained from \eqref{eq:debias_estimate_inverse}.
\end{corollary}

\section{Numerical Experiments}
\label{sec:simulations}  
 To evaluate the finite sample performance of the proposed SAGE estimator, we conduct numerical experiments to: [Aim (i)] assess the bias and variance of the estimates, along with the coverage probability of the confidence intervals; [Aim (ii)] examine how performance varies with sample size; [Aim (iii)] test the feasibility of conducting linear contrast tests; and [Aim (iv)] analyze computation time.

For Aim (i), we simulate $n$ samples $\{(\xb^{(i)}, \ub^{(i)}): i \in [n]\}$ from \eqref{eq:Omega_form} with $n=400$, where $\xb^{(i)} \in \RR^p$ represents $p$ response variables, and $\ub^{(i)} \in \RR^q$ is the external covariate vector, such as SNPs.   The elements $\ub^{(i)}$ are generated independently from a $\Bernoulli(0.5)$ distribution. 
Given $\ub^{(i)}$, 
we set $\Omega(\ub)_{jj} = \sigma^{jj} = 1$, and for the off-diagonal elements, we choose $h = 1, 2$ to allow nonzero values in $\Bb_h$, meaning  the number of effective covariates is 2.
For each $h \in \{1,2\}$ (corresponding to the matrix $\Bb_h$), the values at positions $(j, j+1)$ and $(j+1, j)$ are set to 0.3, with $j = 1$. We also set $\Bb_0=\Ib$. Once $\bOmega(\ub^{(i)})$  are generated, we sample $\xb^{(i)}$ from $\cN(\zero, \bOmega^{-1}(\ub^{(i)}))$ for each $i \in [n]$. For each simulation configuration, we generate 200 independent datasets. 
{Based on the parameter setting, the nonzero coefficients in $\boldsymbol{\beta}$ are located at indexes $\text{ind}_1=p$, $\text{ind}_2=2p - 1$, $\text{ind}_3=(p - 1)(q + 1) + p$ and $\text{ind}_4=(p - 1)(q + 1) + 2p - 1$, respectively.}

As a benchmark for assessing biases, we first compute the pre-debiased multi-task learning estimates \eqref{eq:multi_estimator}  \citep{doi:10.1080/10618600.2024.2421246}.
We then compare two methodologies for bias correction and statistical inference. The first is the SAGE estimator $\hbbeta_j^{u}$ in the multi-task learning framework, with inference results provided in Theorem~\ref{mainthm:Known_Gamma_multi}. {In this methodology as well as for the pre-debiased multi-task  estimates,  we consider two fixed tuning parameter choices, $\lambda_e = 0.3$ and $0.6$, both of order $\sqrt{\log (p)/n}$ (as required by the theorem), along with a cross-validated $\lambda_e$.
 The corresponding $\lambda_g$ is set as $\lambda_e/\sqrt{2}$, based on the ratio $s_e/s_g = 2$.} The second is an oracle method,  serving as a ``golden standard" approach by assuming prior knowledge of the non-zero sets $\cS_j$ and applying ordinary least squares (OLS)  to each non-zero set $\cS_j$ to obtain $\hbbeta_j^{oracle}$ for inference. 
{To ensure comparability with the oracle estimates, we focus on specific sets in the debiasing procedure, such as $\cS$ where $\bbeta_{\cS} \neq 0$. This approach also allows us to evaluate whether the debiased estimates within $\cS$ exhibit a second-order normal distribution centered around the true values. The debiasing parameters are set to $\alpha = 1/\sqrt{n}$ and $\gamma = 2/\sqrt{n}$.}

\begin{figure}[!t]
\centering 
\subfigure[ $(p,q)=(120,20)$]{
\label{fig:X}
\includegraphics[trim=0 2cm 0 0, width=0.48\textwidth]{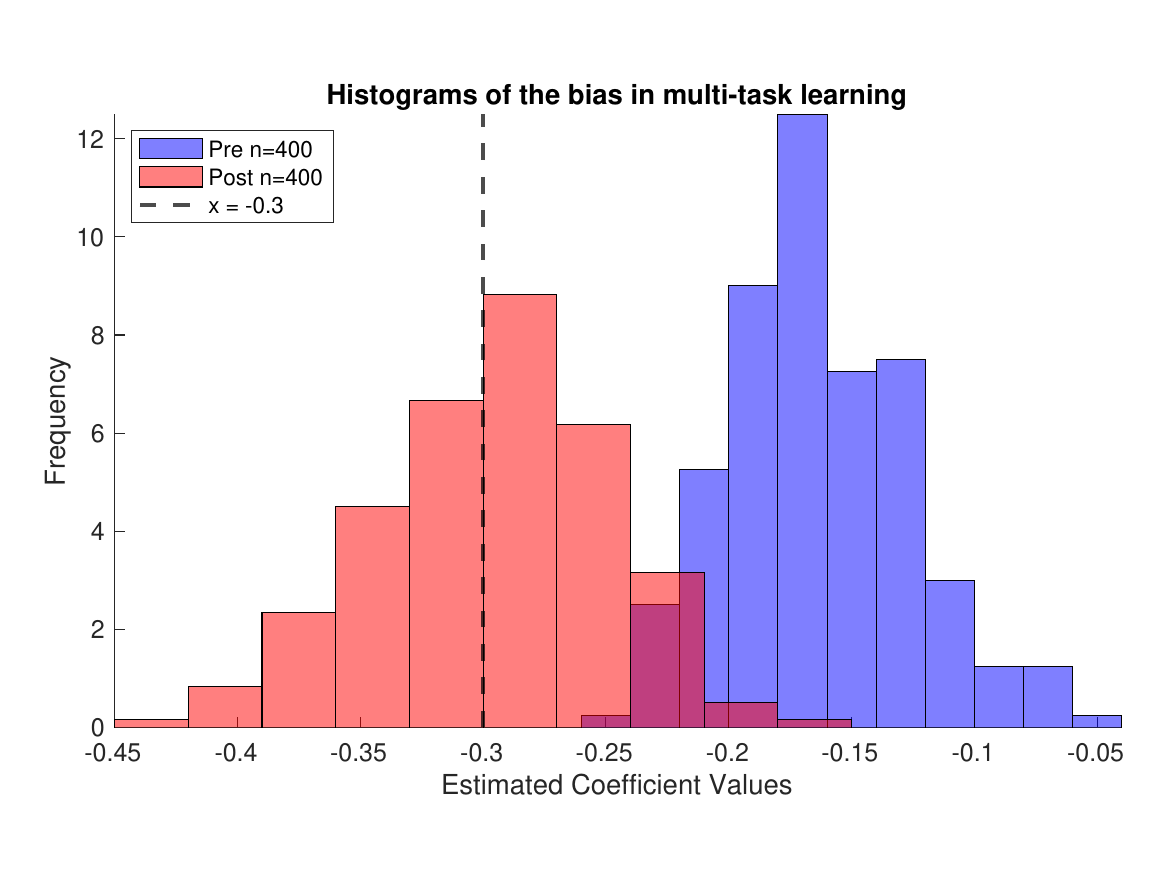}}
\subfigure[ $(p,q)=(20,120)$]{
\label{fig:exp1_2}
\includegraphics[trim=0 2cm 0 0, width=0.48\textwidth]{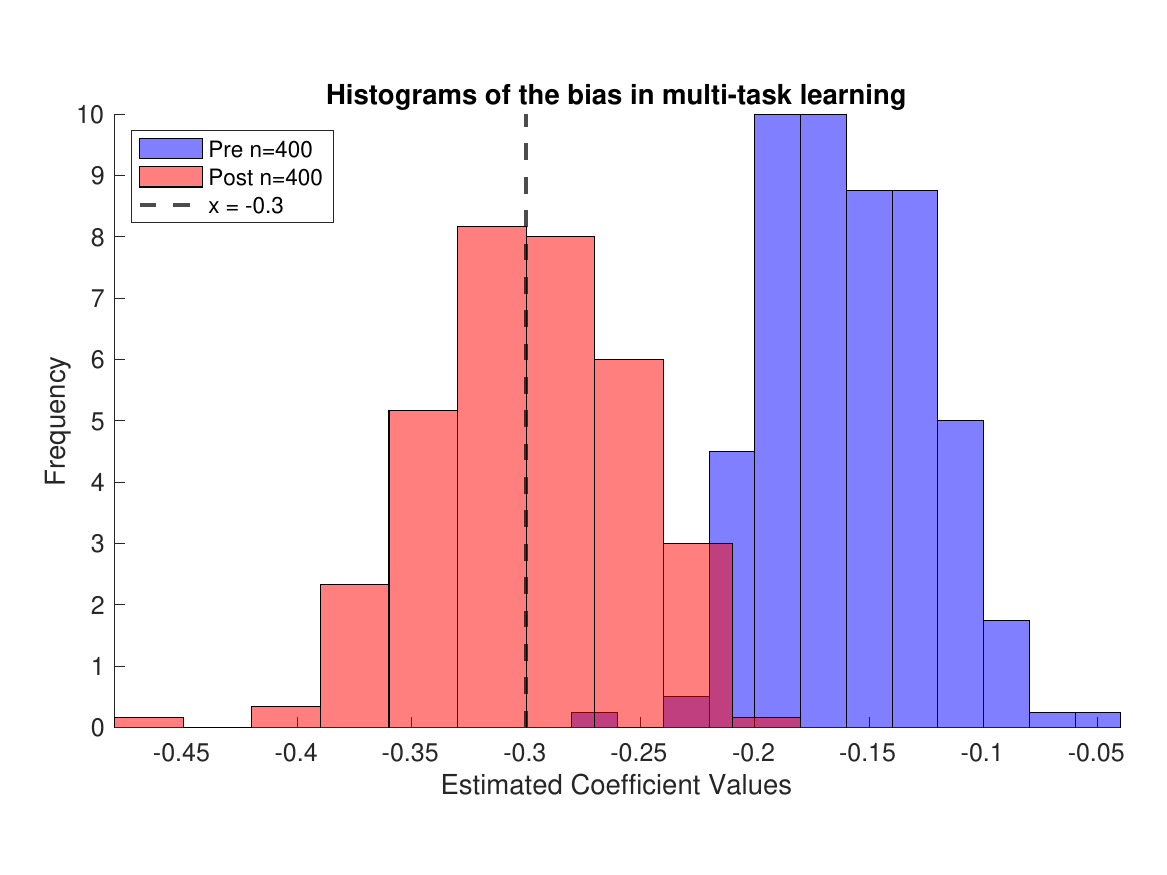}}
\caption{Histograms of {pre-debiased estimates (referred to as Pre) and SAGE estimates (referred to as Post) for $\bbeta_{\text{ind}_1}$} with varying $p$ and $q$.}
\label{fig:n400debiase_histogram}
\end{figure}

We first visually assess the performance of the bias correction of the proposed debiased method by presenting histogram figures of the SAGE 
estimates $\hbbeta^{u}$ across 200 repetitions, plotting the histogram of {$\hbbeta^{u}_{\text{ind}_1}$} 
for $(p,q)=(120,20)$ and  $(20,120)$, respectively. We select the $\hbbeta$ obtained from $(\lambda_e, \lambda_g) = (0.3, 0.212)$ and show in Figure~\ref{fig:n400debiase_histogram} how the SAGE estimates concentrate around the true parameter value of $-0.3$.
As shown in Figure~\ref{fig:n400debiase_histogram}, the undebiased estimates display a clear bias away from the true value of $-0.3$. However, after applying our debiasing procedure, the SAGE 
estimates are tightly concentrated around $-0.3$. The resulting histogram appears nearly normal and symmetric around the line $-0.3$, demonstrating the approximate normality of our proposed estimator. 
We next discuss the performance of statistical inference as reported in Table~\ref{table:n400} {over 200 data replicates}.

\begin{table}[!t]
\centering
{\renewcommand{\baselinestretch}{1.35}
\caption{Simulation results with $n= 400$. {Standard deviations are shown in parentheses}. \textbf{Pre-Bias}:  the average bias of $\hbbeta$ prior to debiasing; \textbf{Post-Bias}: the average bias of the SAGE estimates post debiasing; 
\textbf{Emp-SD}: the empirical standard deviation of debiased estimates after standardization
with the theoretical value being 1;  \textbf{Cov-Prob}: the estimated coverage probability of the 95\% confidence interval.\label{table:n400}}}
\renewcommand{\arraystretch}{0.5}
\begin{tabular}{|ccccccc|}
\hline  $(p,q)$
        &   $(\lambda_e,\lambda_g)$        &      & {$\bbeta_{\text{ind}_1}$}        &  {$\bbeta_{\text{ind}_2}$}      &  {$\bbeta_{\text{ind}_3}$}    &    {$\bbeta_{\text{ind}_4}$}   \\ \hline
\multicolumn{1}{|c|}{\multirow{12}{*}{$(120,20)$}} & \multicolumn{1}{c|}{\multirow{4}{*}{$(.3,.212)$}}      & Pre-Bias  & .137(.036)      & .143(.037)      & .138(.036)      & .143(.038)      \\
\multicolumn{1}{|c|}{}            & \multicolumn{1}{c|}{}               & Post-Bias & .005(.047)      & .011(.046)      & .006(.045)      & .011(.046)      \\
\multicolumn{1}{|c|}{}            & \multicolumn{1}{c|}{}               & Emp-SD    & 1.22      & 1.20      & 1.18      & 1.21      \\
\multicolumn{1}{|c|}{}            & \multicolumn{1}{c|}{}               & Cov-Prob  & 87.5\% & 89.5\% & 89\% & 85.5\% \\  \cline{2-7} 
\multicolumn{1}{|c|}{}            & \multicolumn{1}{c|}{\multirow{4}{*}{$(.6,.424)$}}       & Pre-Bias  & .278(.016)      & .280(.017)      & .278(.016)      & .280(.017)      \\
\multicolumn{1}{|c|}{}            & \multicolumn{1}{c|}{}               & Post-Bias & .013(.033)      & .014(041)      & .013(.032)      & .015(.042)      \\
\multicolumn{1}{|c|}{}            & \multicolumn{1}{c|}{}               & Emp-SD    & 0.87      & 1.08      & 0.83      & 1.11      \\
\multicolumn{1}{|c|}{}            & \multicolumn{1}{c|}{}               & Cov-Prob  & 97\%      & 91.5\%      & 97.5\%      & 90\%      \\\cline{2-7}
\multicolumn{1}{|c|}{}            & \multicolumn{1}{c|}{\multirow{4}{*}{Cross Validation}}       & Pre-Bias  & .116(.048)      & .122(.047)      & .117(.046)      & .122(.047)      \\
\multicolumn{1}{|c|}{}            & \multicolumn{1}{c|}{}               & Post-Bias & .004(.047)      & .010(.046)      & .005(.045)      & .010(.046)      \\
\multicolumn{1}{|c|}{}            & \multicolumn{1}{c|}{}               & Emp-SD    & 1.23      & 1.20      & 1.17      & 1.20      \\
\multicolumn{1}{|c|}{}            & \multicolumn{1}{c|}{}               & Cov-Prob  & 89\%      & 89.5\%      & 89.5\%      & 86.5\%      \\  \hline
\multicolumn{1}{|c|}{\multirow{3}{*}{$(120,20)$}}                    & \multicolumn{1}{c|}{\multirow{3}{*}{$\hbbeta^{oracle}$}} & Bias      & -.007(.046)      & .001(.045)      & -.003(.044)      & .002(.045)      \\
\multicolumn{1}{|c|}{}          & \multicolumn{1}{c|}{}  & Emp-SD    & 0.91      & 0.89      & 0.88      & 0.90      \\
\multicolumn{1}{|c|}{}          & \multicolumn{1}{c|}{}  & Cov-Prob  & 98\%      & 98.5\%      & 97.5\%      & 99\%      \\  \hline
\multicolumn{1}{|c|}{\multirow{12}{*}{$(20,120)$}} & \multicolumn{1}{c|}{\multirow{4}{*}{$(.3,.212)$}}       & Pre-Bias  & .142(.035)      & .146(.040)      & .143(.035)      & .148(.040)      \\
\multicolumn{1}{|c|}{}            & \multicolumn{1}{c|}{}               & Post-Bias  & .002(.044)      & .006(.049)      & .006(.043)      & .010(.047)       \\
\multicolumn{1}{|c|}{}            & \multicolumn{1}{c|}{}               & Emp-SD    & 1.11      & 1.23      & 1.09      & 1.20      \\
\multicolumn{1}{|c|}{}            & \multicolumn{1}{c|}{}               & Cov-Prob  & 94\%      & 90\%      & 93.5\%      & 88.5\% \\ \cline{2-7} 
\multicolumn{1}{|c|}{}            & \multicolumn{1}{c|}{\multirow{4}{*}{$(.6,.424)$}}       & Pre-Bias  & .279(.017)      & .279(.017)       & .279(.017)      & .279(.016)      \\
\multicolumn{1}{|c|}{}            & \multicolumn{1}{c|}{}               & Post-Bias & .009(.035)      & .009(.045)      & .013(.034)      & .012(.044)      \\
\multicolumn{1}{|c|}{}            & \multicolumn{1}{c|}{}               & Emp-SD    & 0.89      & 1.14      & 0.85      & 1.12      \\
\multicolumn{1}{|c|}{}            & \multicolumn{1}{c|}{}               & Cov-Prob  & 96.5\%      & 89.5\%      & 96.5\%      & 92\%      \\\cline{2-7}
\multicolumn{1}{|c|}{}            & \multicolumn{1}{c|}{\multirow{4}{*}{Cross Validation}}        & Pre-Bias  & .104(.043)      & .108(.048)      & .107(.043)      & .111(.048)      \\
\multicolumn{1}{|c|}{}            & \multicolumn{1}{c|}{}               & Post-Bias & .004(.044)      & .008(.048)      & .008(.043)      & .012(.048)      \\
\multicolumn{1}{|c|}{}            & \multicolumn{1}{c|}{}               & Emp-SD    & 1.11      & 1.22      & 1.10      & 1.21      \\
\multicolumn{1}{|c|}{}            & \multicolumn{1}{c|}{}               & Cov-Prob  & 93.5\%      & 90.5\%      & 93\%      & 87.5\%      \\ \hline
\multicolumn{1}{|c|}{\multirow{3}{*}{$(20,120)$}}                    & \multicolumn{1}{c|}{\multirow{3}{*}{$\hbbeta^{oracle}$}} & Bias      & -.003(.043)      & .001(.047)      & .002(.043)      & .006(.046)      \\
\multicolumn{1}{|c|}{}          & \multicolumn{1}{c|}{}  & Emp-SD    & 0.86      & 0.94      & 0.86      & 0.91      \\
\multicolumn{1}{|c|}{}          & \multicolumn{1}{c|}{}  & Cov-Prob  & 98.5\%      & 95.5\%      & 97.5\%      & 96\%      \\  \hline
\end{tabular}
\end{table}

The tabulated results confirm that the proposed SAGE estimator consistently outperforms the undebiased estimator across all four indices. The undebiased estimates show significantly larger pre-bias values, reflecting systematic bias, while the SAGE estimates achieve substantially reduced post-bias values, offering more accurate parameter estimates. Our SAGE estimator also performs comparably to the oracle estimator, which achieves minimal bias and near-nominal coverage probability. This highlights the robustness of the SAGE estimator in providing reliable inference without knowledge of the true support. Additionally, the theoretical variance aligns well with empirical variance, though variations in tuning parameters can affect this alignment. For instance, smaller tuning parameters may lead to model-based variance overestimating true variance, causing slight under-coverage in confidence intervals, likely due to finite sample effects.

\begin{table}[!t]
\centering
\renewcommand{\arraystretch}{0.5}
\caption{Simulation results with $n= 800$.}
\label{table:n800}
\begin{tabular}{|ccccccc|}
\hline  $(p,q)$
        &   $(\lambda_e,\lambda_g)$        &    & {$\bbeta_{\text{ind}_1}$}        &   {$\bbeta_{\text{ind}_2}$}      &  {$\bbeta_{\text{ind}_3}$}     &    {$\bbeta_{\text{ind}_4}$}   \\ \hline
\multicolumn{1}{|c|}{\multirow{4}{*}{$(120,20)$}} & \multicolumn{1}{c|}{\multirow{4}{*}{$(.3,.212)$}}      & Pre-Bias  & .144(.029)      & .139(.026)      & .144(.028)      & .139(.027)      \\
\multicolumn{1}{|c|}{}            & \multicolumn{1}{c|}{}               & Post-Bias & 005(.034)      & -.001(.031)      & .006(.033)      & .001(.032)      \\
\multicolumn{1}{|c|}{}            & \multicolumn{1}{c|}{}               & Emp-SD    & 1.21      & 1.10      & 1.17      & 1.16      \\
\multicolumn{1}{|c|}{}            & \multicolumn{1}{c|}{}               & Cov-Prob  & 91\% & 91.5\% & 90\% & 91\% \\  \hline
\multicolumn{1}{|c|}{\multirow{4}{*}{$(20,120)$}} & \multicolumn{1}{c|}{\multirow{4}{*}{$(.3,.212)$}}      & Pre-Bias  & .141(.025)      & .140(.028)      & .143(.026)      & .141(.027)      \\
\multicolumn{1}{|c|}{}            & \multicolumn{1}{c|}{}               & Post-Bias & .001(.030)      & -.000(.034)      & .005(.031)      & .003(.032)      \\
\multicolumn{1}{|c|}{}            & \multicolumn{1}{c|}{}               & Emp-SD    & 1.07      & 1.18      & 1.11      & 1.13      \\
\multicolumn{1}{|c|}{}            & \multicolumn{1}{c|}{}               & Cov-Prob  & 93\% & 87.5\% & 92\% & 90.5\% \\  \hline
\end{tabular}
\end{table}


{To confirm this and for Aim (ii), we conduct more experiments to examine the performance by varying  $n$ from 100 to 800.}  The results for $(\lambda_e,\lambda_g)=(0.3,0.212)$ {over 200 data replications} are reported in Table~\ref{table:n800} and additional results are in Section~\ref{sec:additional_simulation} of the supplement.
These results show that, when the sample size is small, the under-coverage of confidence intervals becomes more pronounced. However, as the sample size increases, such as $n=800$, the empirical variance aligns more closely with the theoretical values and the bias of the SAGE estimates becomes closer to $0$.  

For Aim (iii),  we  focus on the case of $j=1$,    where the non-zero indices for $\bbeta_1$ are $p$ and $2p-1$. We estimate the following linear combinations and conduct testing for the corresponding linear contrasts respectively:
$$
\text{I}:(\bbeta_1)_p-(\bbeta_1)_{2p-1};\quad\text{II}:(\bbeta_1)_1-(\bbeta_1)_{p};\quad\text{III}:(\bbeta_1)_1+2(\bbeta_1)_{2};\quad
\text{IV}:\begin{pmatrix}
    2(\bbeta_1)_1-(\bbeta_1)_p\\
    (\bbeta_1)_2+(\bbeta_1)_{p}
\end{pmatrix}.
$$
Case I examines the relationship between two non-zero parameters, Case II compares a non-zero parameter with a zero parameter, Case III assesses the relationship between two zero parameters, and Case IV explores simultaneous inference. We set $(\lambda_e, \lambda_g) = (0.6, 0.424)$. Based on Corollary~\ref{coro:2}, the non-zero rows of matrix $\Ab$ corresponding to the four cases is given by $(\eb_p-\eb_{2p-1})^\top$, $(\eb_1-\eb_{2p-1})^\top$, $(\eb_1+2\eb_2)^\top$ and 
$\genfrac{(}{)}{0pt}{}{(2\eb_1-\eb_p)^\top}{(\eb_2+\eb_{p})^\top}$,
and the values of $\Ab\bbeta$ for the four cases are $0,0.3,0$ and $(0.3,-0.3)^\top$. 

\begin{table}[!t]
\centering
{\renewcommand{\baselinestretch}{1.35}\caption{Results for the four cases with $n=400$. \textbf{Emp-AVE}: empirical mean of SAGE estimates after standardization
with the theoretical value being 0; \textbf{Emp-SD}:  empirical standard deviation of SAGE  estimates after standardization
with the theoretical value being 1;  \textbf{Cov-Prob}:  estimated coverage probability of the 95\% confidence interval. V1 and V2 correspond to the two vector values being tested in Case IV.}
\label{table:testresults}}
\renewcommand{\arraystretch}{0.5}
\begin{tabular}{|ccccccc|}
\hline
\multirow{2}{*}{$(p,q)$}      & & \multirow{2}{*}{I} & \multirow{2}{*}{II} & \multirow{2}{*}{III} & \multicolumn{2}{c|}{IV} \\
  & &&&& V1        & V2       \\ \hline
\multicolumn{1}{|c|}{\multirow{3}{*}{$(120,20)$}} & Emp-Ave                     & -.103                        & -.182                        & .009                        & -.128            & .146           \\
\multicolumn{1}{|c|}{}        & Emp-SD                      & 1.15                        & 1.14                        & 1.10                        & 1.17            & 1.05           \\
\multicolumn{1}{|c|}{}        & Cov-Prob                    & 92.5\%                        & 92\%                        & 92\%                        & \multicolumn{2}{c|}{91.5\%}       \\ \hline
\multicolumn{1}{|c|}{\multirow{3}{*}{$(20,120)$}}  & Emp-Ave                    & -.100                        & -.125                        & .069                        & -.086            & .151           \\
\multicolumn{1}{|c|}{}        & Emp-SD                      & 1.09                        & 1.12                        & 1.11                        & 1.18            & 1.14            \\
\multicolumn{1}{|c|}{}        & Cov-Prob                    & 93\%                        & 92.5\%                        & 94.5\%                        & \multicolumn{2}{c|}{89.5\%}       \\ \hline
\end{tabular}
\end{table}

Table~\ref{table:testresults} presents the results
based on  200 repetitions.   
As shown in Table~\ref{table:testresults},  the results of  Corollary~\ref{coro:2} hold under the examined  hypotheses. In particular, under Case IV which involves two different dimensions  and shares the same $(\bbeta_1)_p$ in both values, the standardized bivariate SAGE estimates, i.e., $\sqrt{n}(\Ab\hMb_j^\top\hbSigma_{\Wb_j}\hMb_j\Ab^\top)^{-1/2}(\Ab\hbbeta_1-\Ab\bbeta_1)$,   approximately follows  a two-dimensional normal distribution with mean 0 and an identity covariance matrix. The Q-Q plots in Figure~\ref{fig:n400qqplot} further confirm the asymptotic standard normal distribution. The joint distribution shown in Figure~\ref{fig:sub_qqplot4} shows a small Pearson correlation of only 0.057 between the two dimensions, and the histograms for both variables exhibit a shape consistent with the standard normal distribution.
\begin{figure}[!t]
\centering 
\subfigure[Results in Case I]{
\label{fig:sub_qqplot1}
\includegraphics[trim=0 1cm 0 0, width=0.4\textwidth]{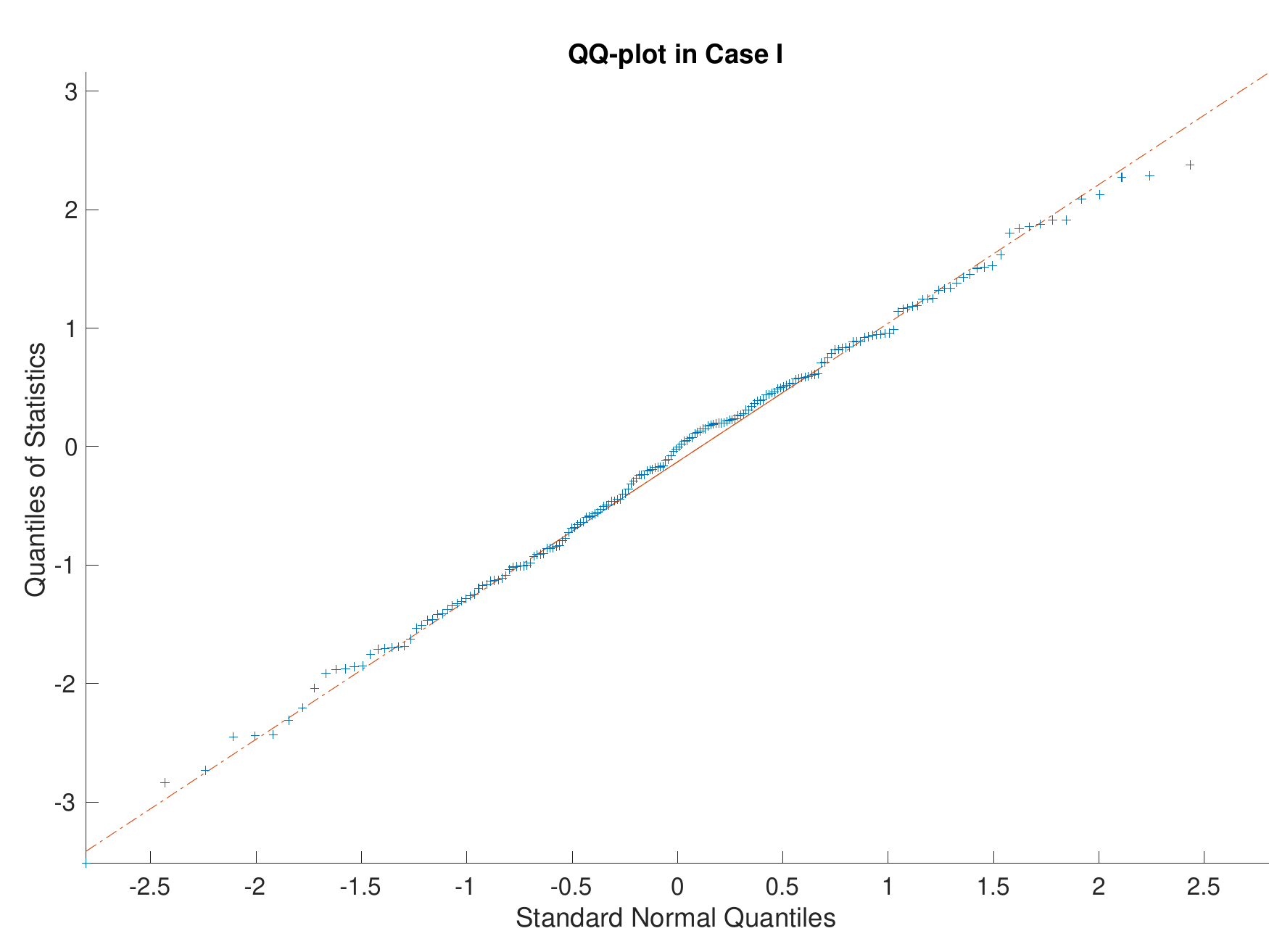}}
\subfigure[Results in Case II]{
\label{fig:sub_qqplot2}
\includegraphics[trim=0 1cm 0 0, width=0.4\textwidth]{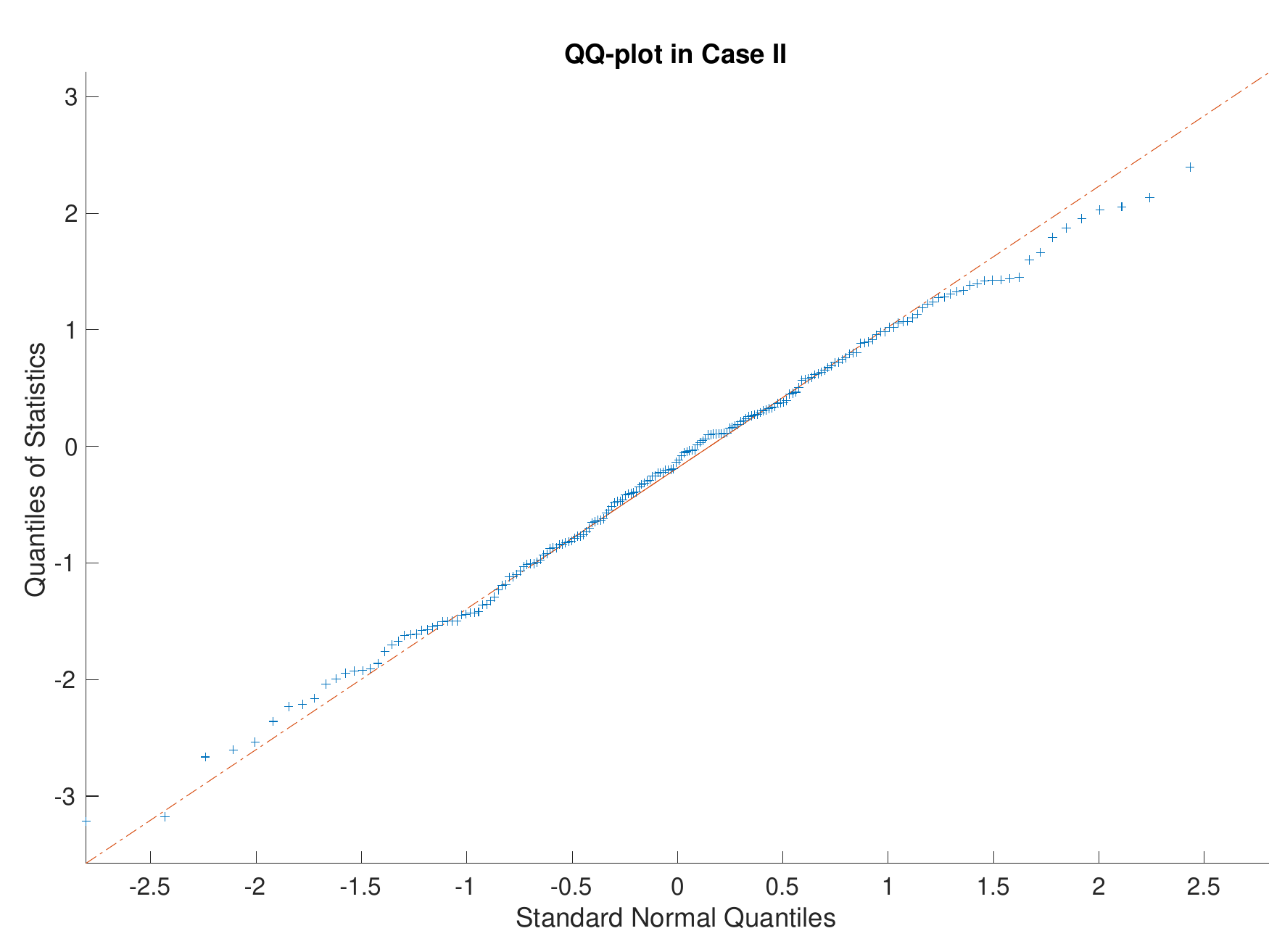}}

\subfigure[Results in Case III]{
\label{fig:sub_qqplot3}
\includegraphics[trim=0 1cm 0 0, width=0.4\textwidth]{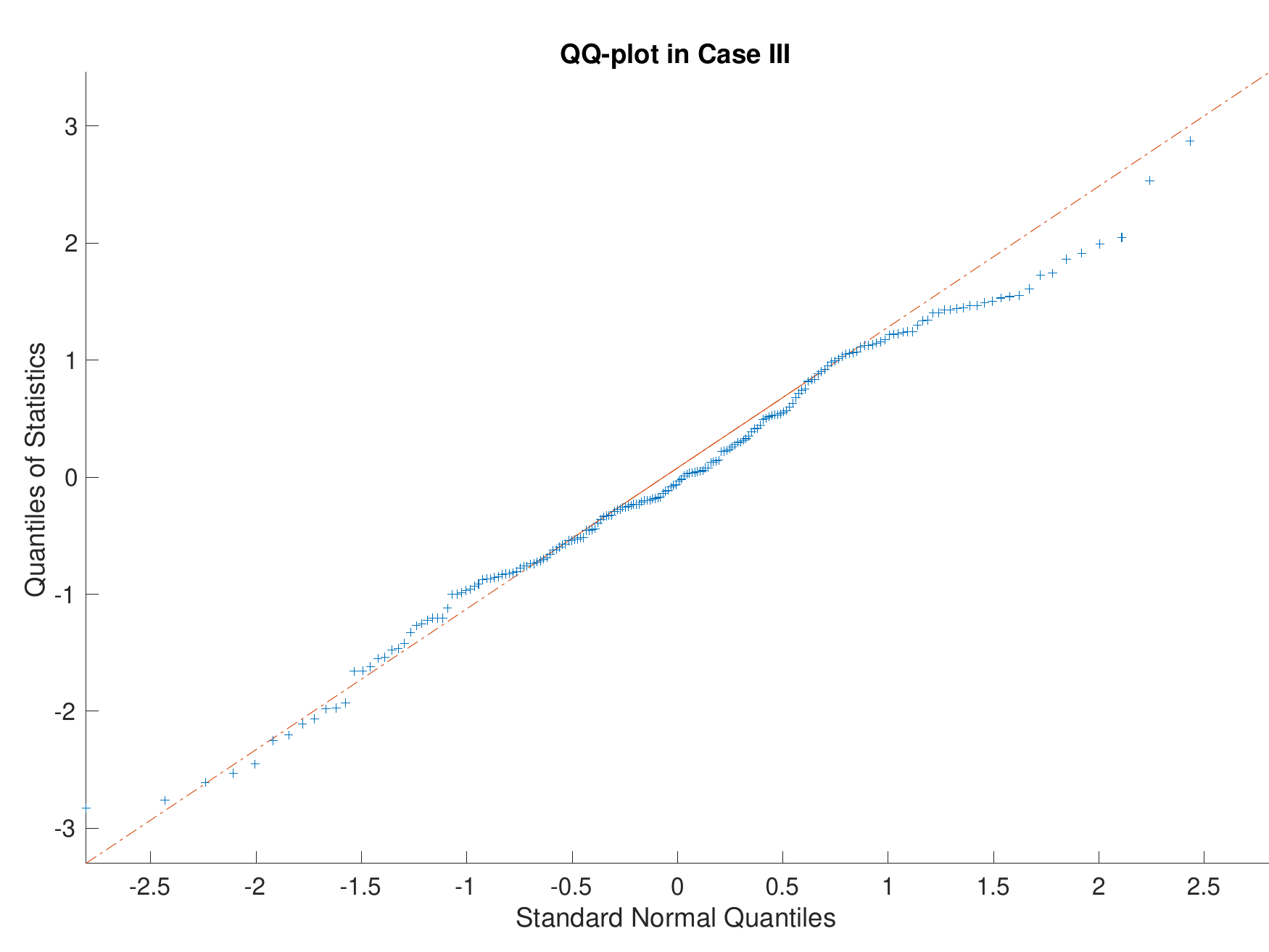}}
\subfigure[Results in Case IV]{
\label{fig:sub_qqplot4}
\includegraphics[trim=0 1cm 0 0, width=0.4\textwidth]{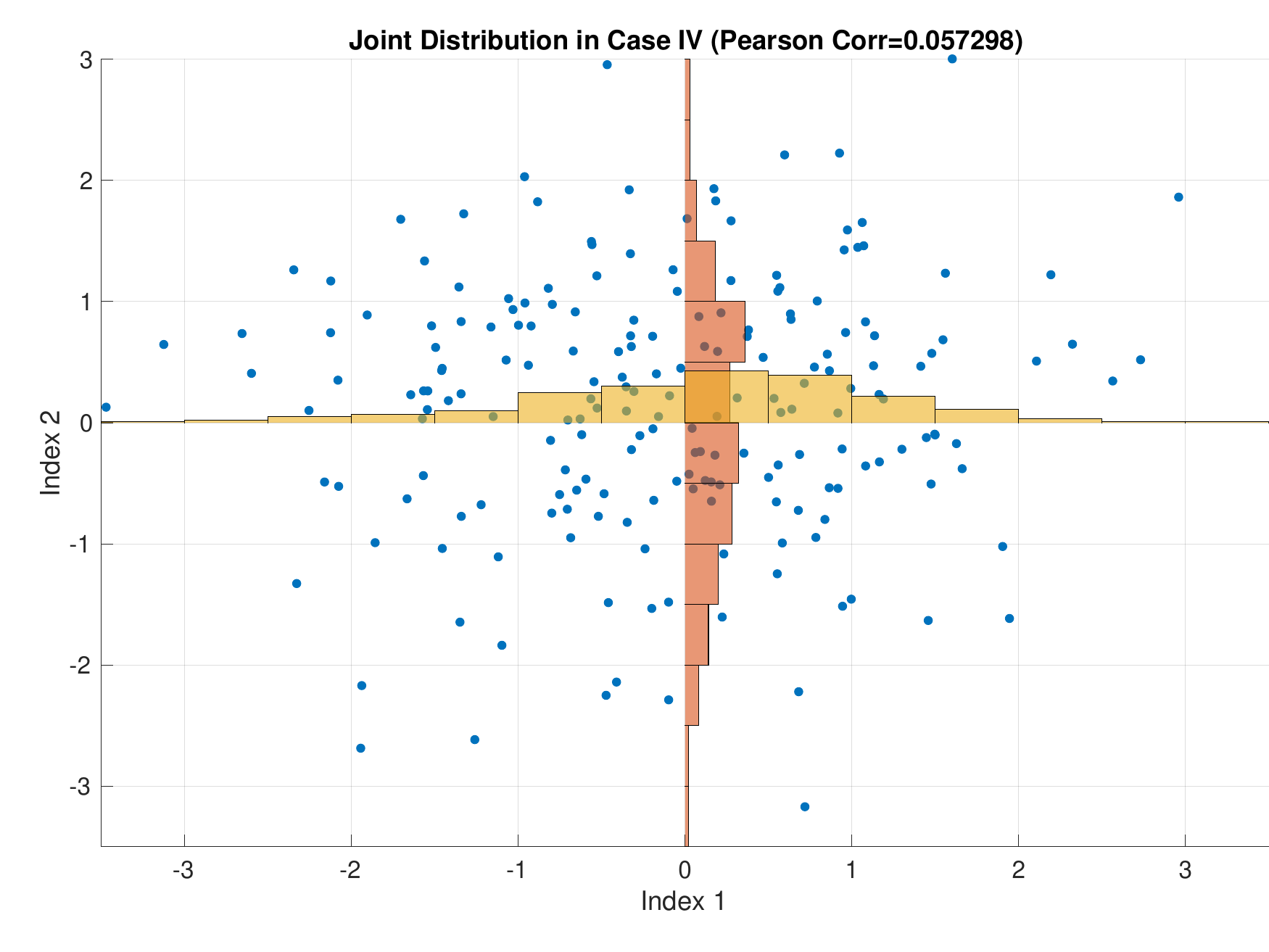}}
{\renewcommand{\baselinestretch}{1.35}
\caption{The example figures demonstrate the asymptotic standard normal distribution behavior of  the standardized SAGE estimates,
$\sqrt{n}(\Ab\hMb_j^\top\hbSigma_{\Wb_j}\hMb_j\Ab^\top)^{-1/2}(\Ab\hbbeta_1-\Ab\bbeta_1)$, in the cases with $(p,q)=(120,20)$. Figures~\ref{fig:sub_qqplot1}, \ref{fig:sub_qqplot2} and \ref{fig:sub_qqplot3} present QQ plots for the Case I, Case II and Case III, respectively, illustrating the asymptotic normality. Figure \ref{fig:sub_qqplot4} shows the joint distribution of two asymptotically independent standard normal variables under Case IV.\label{fig:n400qqplot}}}
\end{figure}

Finally, for Aim (iv), we  conduct an experiment on our debiasing algorithm in \eqref{eq:debias_estimate_inverse}, comparing it to the optimization problem in \eqref{eq:debias_estimate_inverse_m} similar to previous works \citep{javanmard2014confidence,cai2022sparse}, to evaluate the computation time. For simplicity, we focus on solving  \eqref{eq:debias_estimate_inverse_m} and \eqref{eq:debias_estimate_inverse}, where we set $l=1$ for the comparison of computation time. 
The simulations run on a MAC Pro with M3 Pro chips. 
Table~\ref{table:computation} compares the computation time between our debiasing procedure by \eqref{eq:debias_estimate_inverse} and the direct optimization of \eqref{eq:debias_estimate_inverse_m}. 
The results clearly indicate that the projection method is significantly faster. As $p$ increases, the proposed projection method remains feasible, while the direct optimization of \eqref{eq:debias_estimate_inverse_m} becomes increasingly computationally prohibitive.

 
\begin{table}[!t]
\centering
\caption{Computation time in seconds with varying $n,p$ while $q=20$.}
\label{table:computation}
\setlength{\tabcolsep}{4pt}
\renewcommand{\arraystretch}{0.65}
\begin{tabular}{|c|ccc|ccc|ccc|}
\hline \multirow{2}{*}{Time ($s$)}
& \multicolumn{3}{c|}{$n=50$} & \multicolumn{3}{c|}{$n=100$} & \multicolumn{3}{c|}{$n=200$} \\\cline{2-10}
& $p=20$   & $p=50$   & $p=100$   & $p=20$    & $p=50$   & $p=100$   & $p=20$    & $p=50$   & $p=100$   \\ \hline
\eqref{eq:debias_estimate_inverse_m}     & 1.36    & 15.76    & 99.14     & 1.55     & 17.52    & 117.92     & 1.25    & 12.56    & 98.38     \\ 
\eqref{eq:debias_estimate_inverse}   & 0.11    & 0.10    & 0.13     & 0.16    & 0.17    & 0.29      & 0.26      & 0.57    & 0.67      \\\hline
\end{tabular}
\end{table}

\section{Analysis of glioblastoma multiforme gene expression graphs}
\label{sec:real_data}
Glioblastoma multiforme is a lethal brain cancer, and existing therapies are largely ineffective \citep{kwiatkowska2013strategies}. To develop effective treatments, such as novel gene therapies, a better understanding of the disease's molecular mechanisms is critical. We apply our methods to infer the effects of single nucleotide polymorphisms (SNPs) on gene co-expression in a Glioblastoma  trial. The dataset, publicly  available via the NIH  Gene Expression Omnibus  database (labeled as GSE108476),  comprises $n=178$ glioblastoma multiforme patients with both microarray and SNP chip profiling.
 
 After preprocessing the raw data as outlined by \citet{gusev2018rembrandt},
we investigate the expression levels of 73 genes associated with the human glioma pathway, as recorded in the Kyoto Encyclopedia of Genes and Genomes (KEGG) database \citep{kanehisa2000kegg}. The covariates include SNPs located within 2kb upstream and 0.5kb downstream of these genes, resulting in 118 nearby SNPs. These SNPs are encoded as ``0'' for the common genotype and ``1'' for all other genotypes. Age and gender are also included as covariates. In total, there are 120 covariates, leading to 317,988  parameters ($73 \times 36 \times 121$). We perform Gaussian graphical regression on age, gender, and SNPs to examine their influence on the graph structure. When fitting the model, we tune the parameters $(\lambda_e,\lambda_g)$
via cross-validation as done in simulations,
and then assess whether the initially dense edges were truly significant by applying the proposed debiased method, i.e.,  the SAGE estimator, for inference. 
We set $\alpha = 1/\sqrt{178}$ and $\gamma = 2/\sqrt{178}$ respectively as guided by Theorem \ref{mainthm:Known_Gamma_multi}.
 
\begin{figure}[htbp]
\centering
\subfigure[Original Selected Edges]{
\label{fig:realh0-origin}
\includegraphics[width=0.31\textwidth]{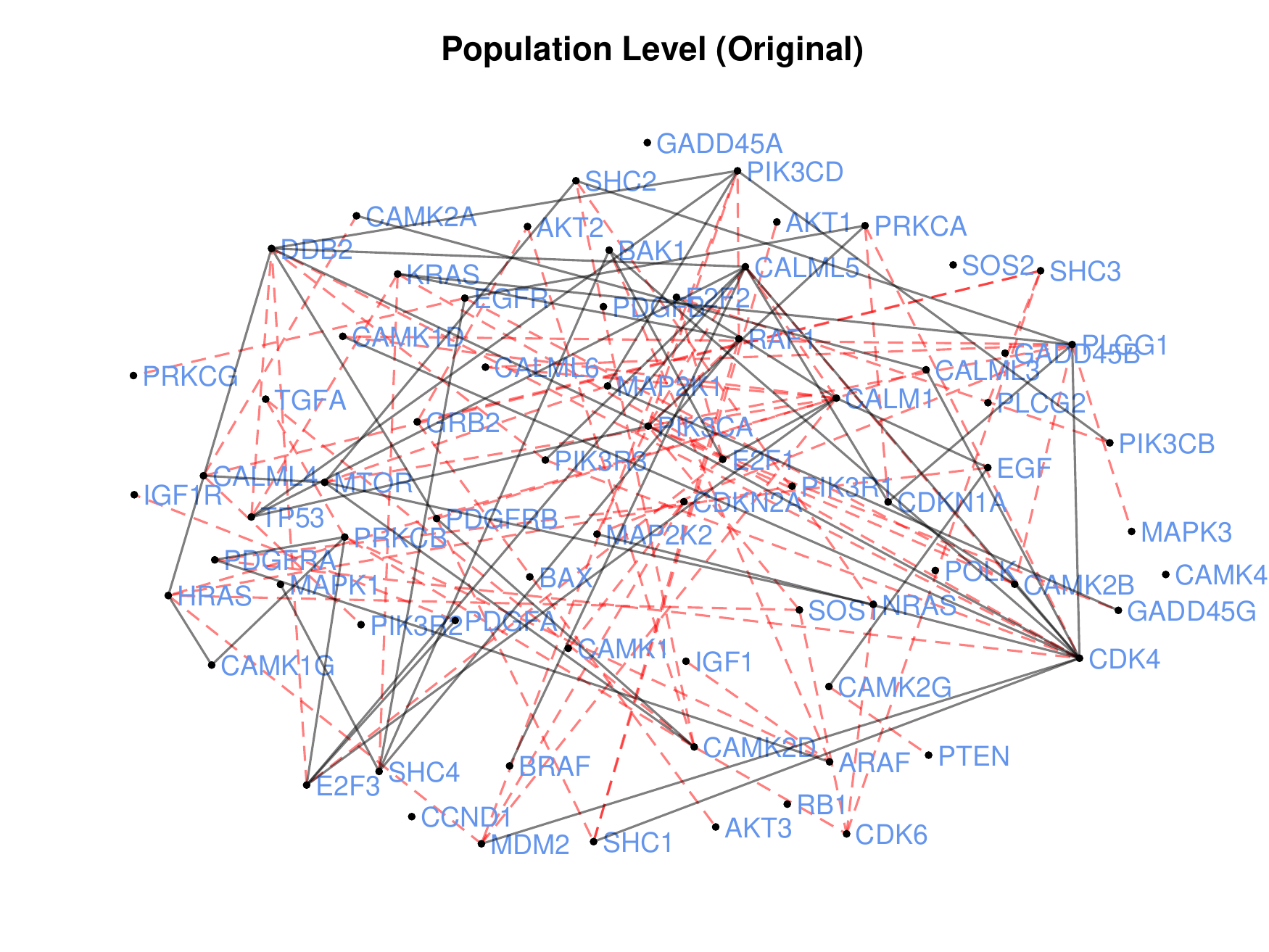}}
\subfigure[Significance Level 0.05]{
\label{fig:realh0-005}
\includegraphics[width=0.31\textwidth]{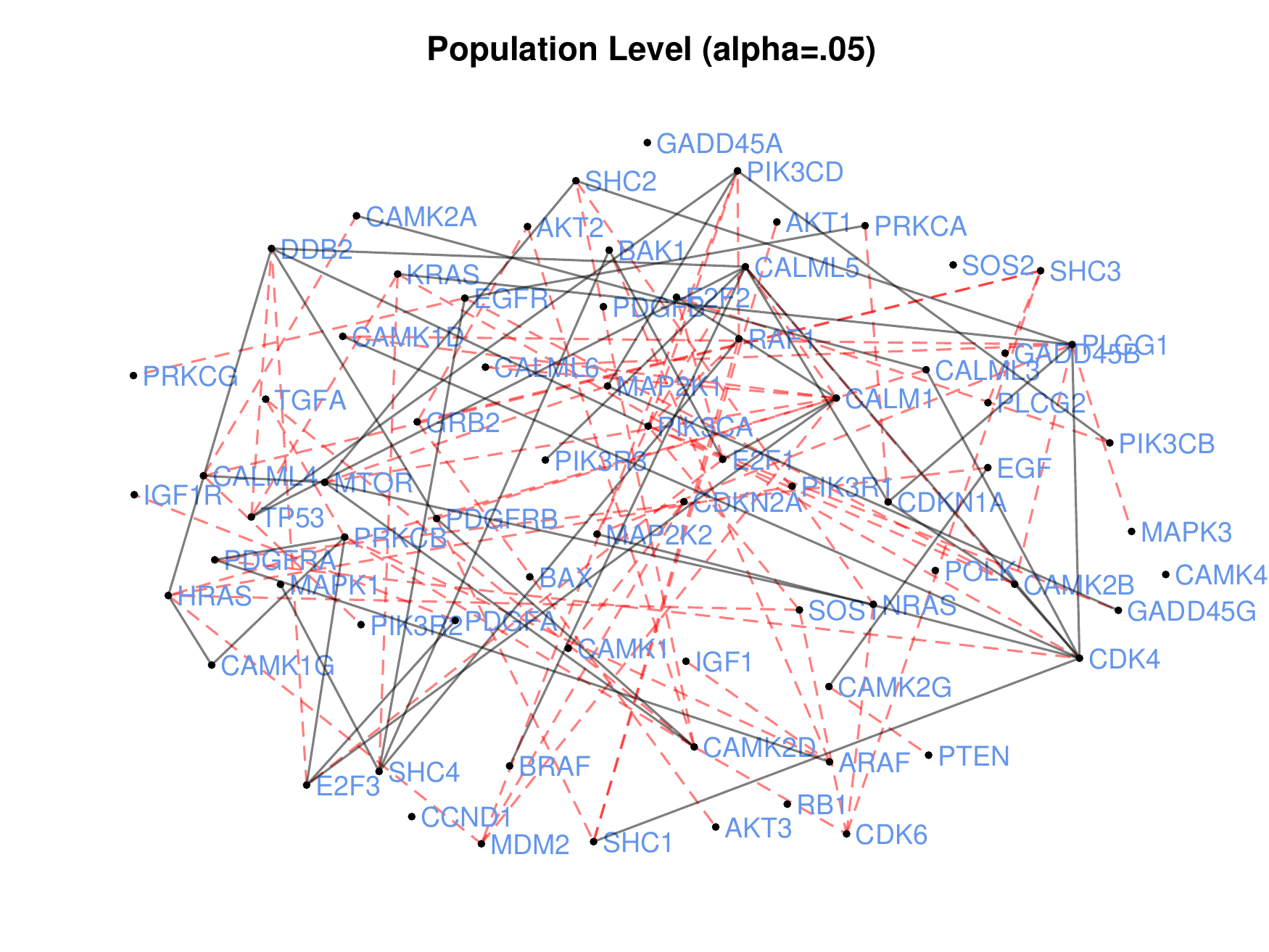}}
\subfigure[Significance Level 0.001]{
\label{fig:realh0-0001}
\includegraphics[width=0.31\textwidth]{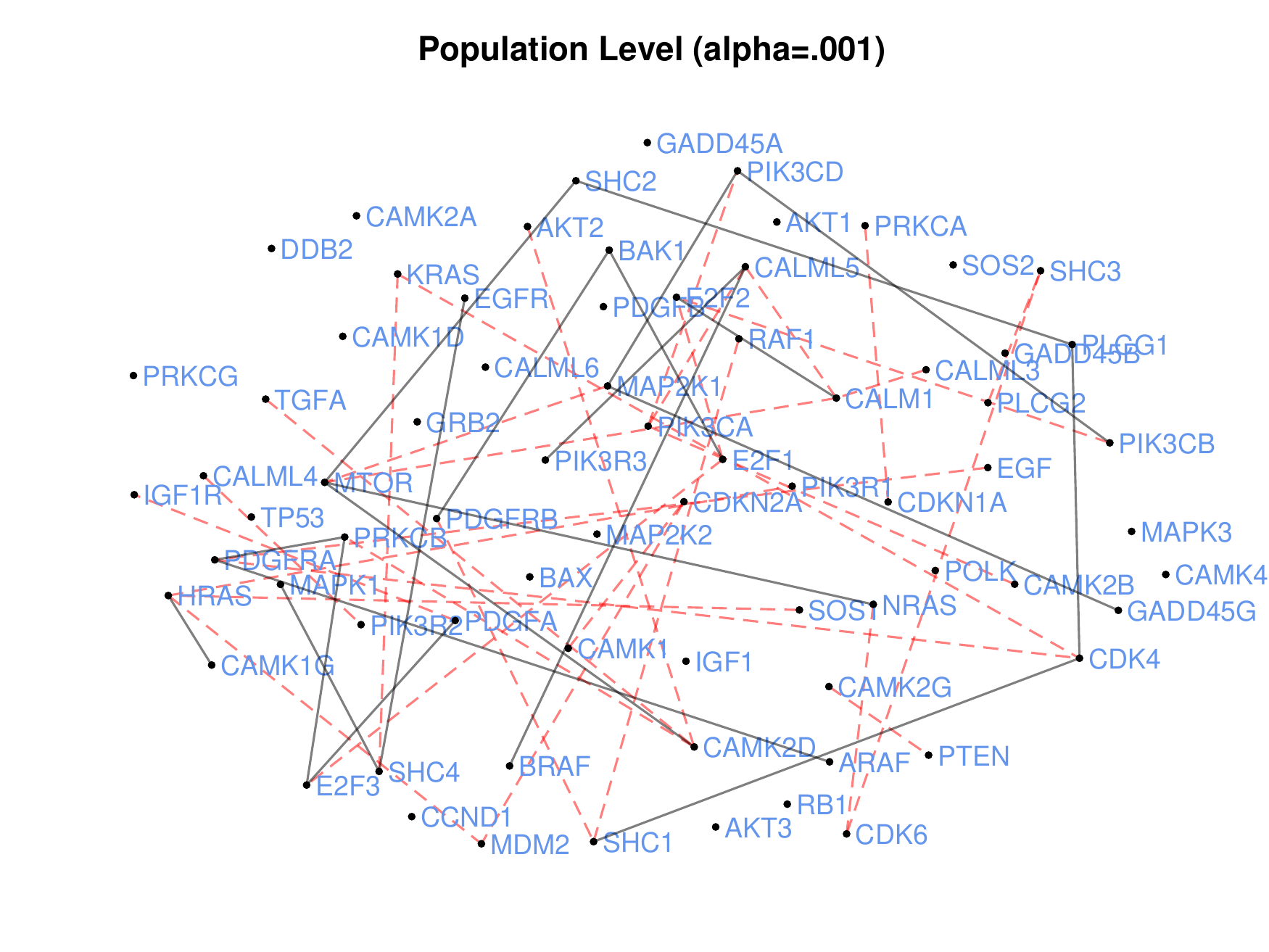}}
{\renewcommand{\baselinestretch}{1.35}\caption{Population-level gene co-expression graph (left), shown with significance levels of 0.05 (middle) and 0.001 (right). Positive partial correlations are shown with red dashed lines, while negative correlations are indicated by black solid lines.}
\label{fig:realgraphpopulation}}
\end{figure}

We first establish the population-level gene network based on the multi-task learning estimates (prior to debiasing), as shown in Figure~\ref{fig:realh0-origin}. We then perform debiasing and test the statistical significance of the detected edges using our inference methodology. Figures~\ref{fig:realh0-005} and \ref{fig:realh0-0001}  demonstrate that after debiasing and statistical testing, the number of detected edges was significantly reduced, retaining only those with strong statistical evidence.  Our results have  biological implications. For instance, the PI3K/AKT/MTOR pathway  plays  a vital role  in glioblastoma and other malignancies \citep{samuels2004oncogenic, cancer2008comprehensive}, while  EGFR, a major oncogene often mutated or amplified in glioblastoma, promotes tumor growth and progression via this  pathway \citep{ekstrand1991genes, melenhorst2008epidermal}. Although EGFR exhibits limited direct connections at a significance level of 0.001, its interaction with SHC4, an intermediary that connects to additional genes, suggests a broader influence throughout the graph. Furthermore, at a more relaxed significance level of 0.05, EGFR shows connections to numerous genes, further highlighting its impact within the network.  The finding regarding EGFR is supported by other literature that has identified its central role in glioblastoma and various other cancers \citep{samuels2004oncogenic, cancer2008comprehensive, ekstrand1991genes}. Moreover, as shown in Figure~\ref{fig:realh0-0001}, our findings align with those of \citet{doi:10.1080/10618600.2024.2421246}, reinforcing the importance of targeting  interconnected pathways in glioblastoma treatment strategies.

\begin{figure}[!t]
\vspace{-14pt}
\centering 
\subfigure[Original Selected Edges]{
\label{fig:realh19-origin}
\includegraphics[width=0.31\textwidth]{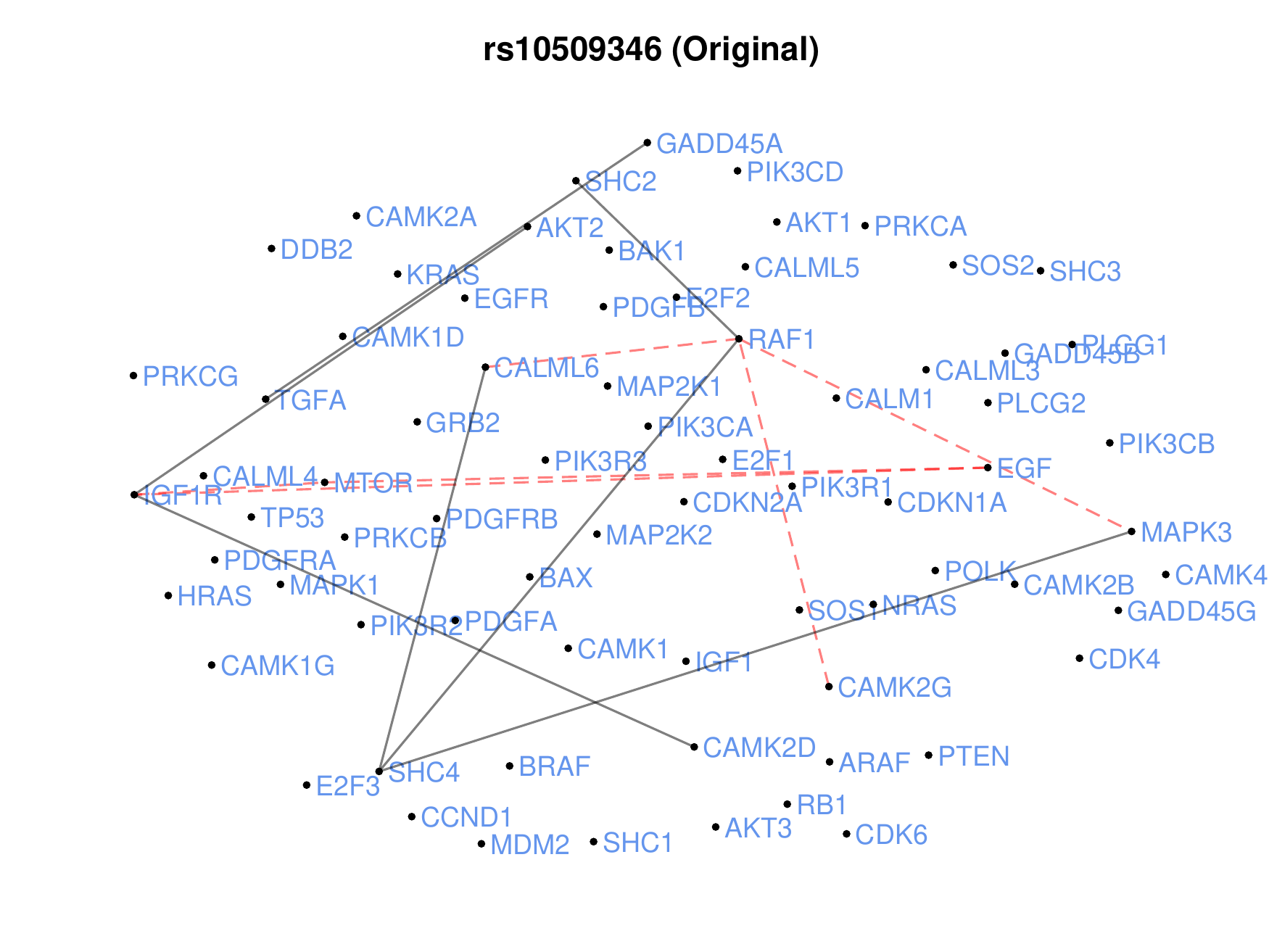}}
\subfigure[Significance Level 0.05]{
\label{fig:realh19-005}
\includegraphics[width=0.31\textwidth]{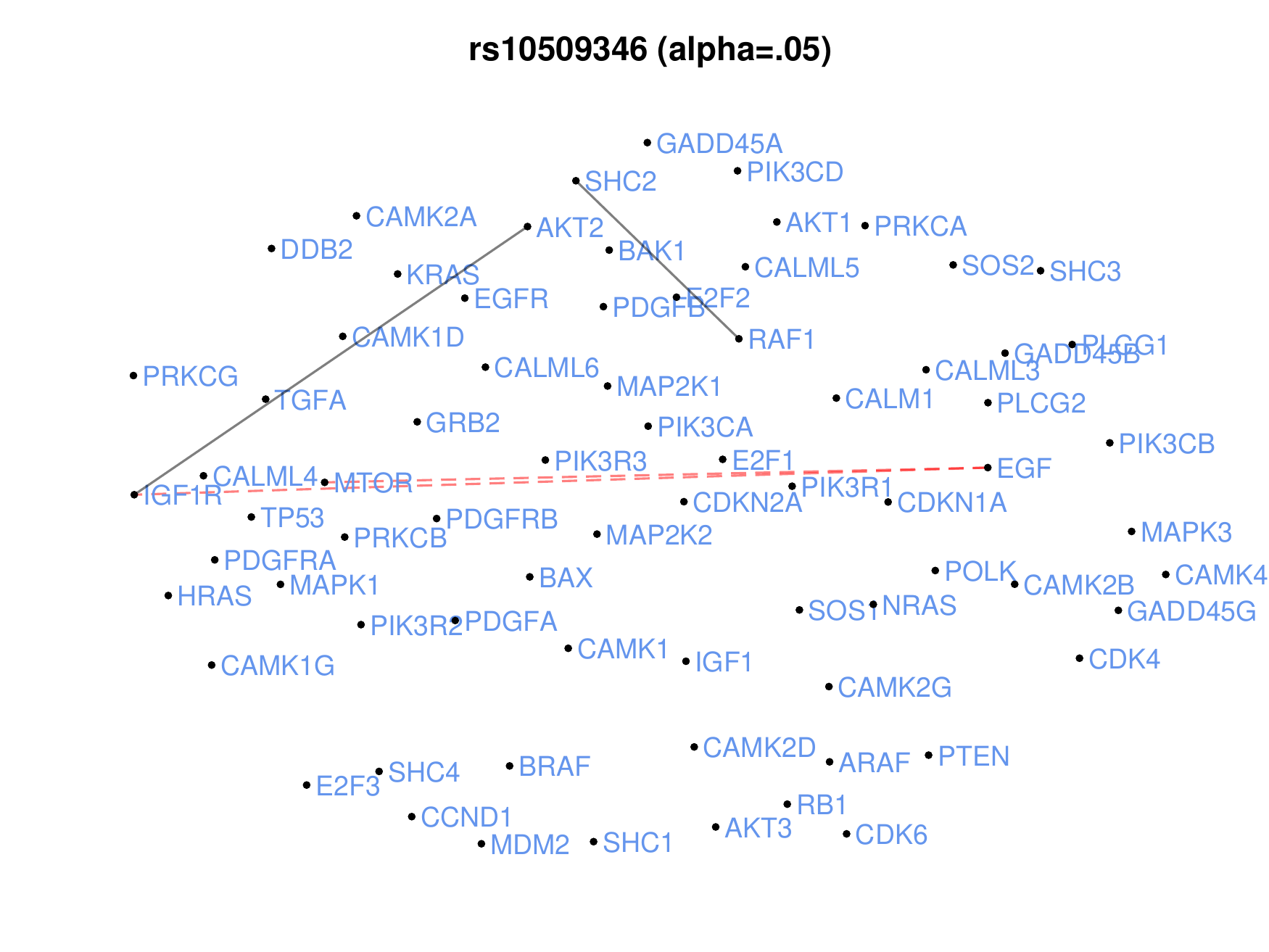}}
\subfigure[Significance Level 0.001]{
\label{fig:realh19-0001}
\includegraphics[width=0.31\textwidth]{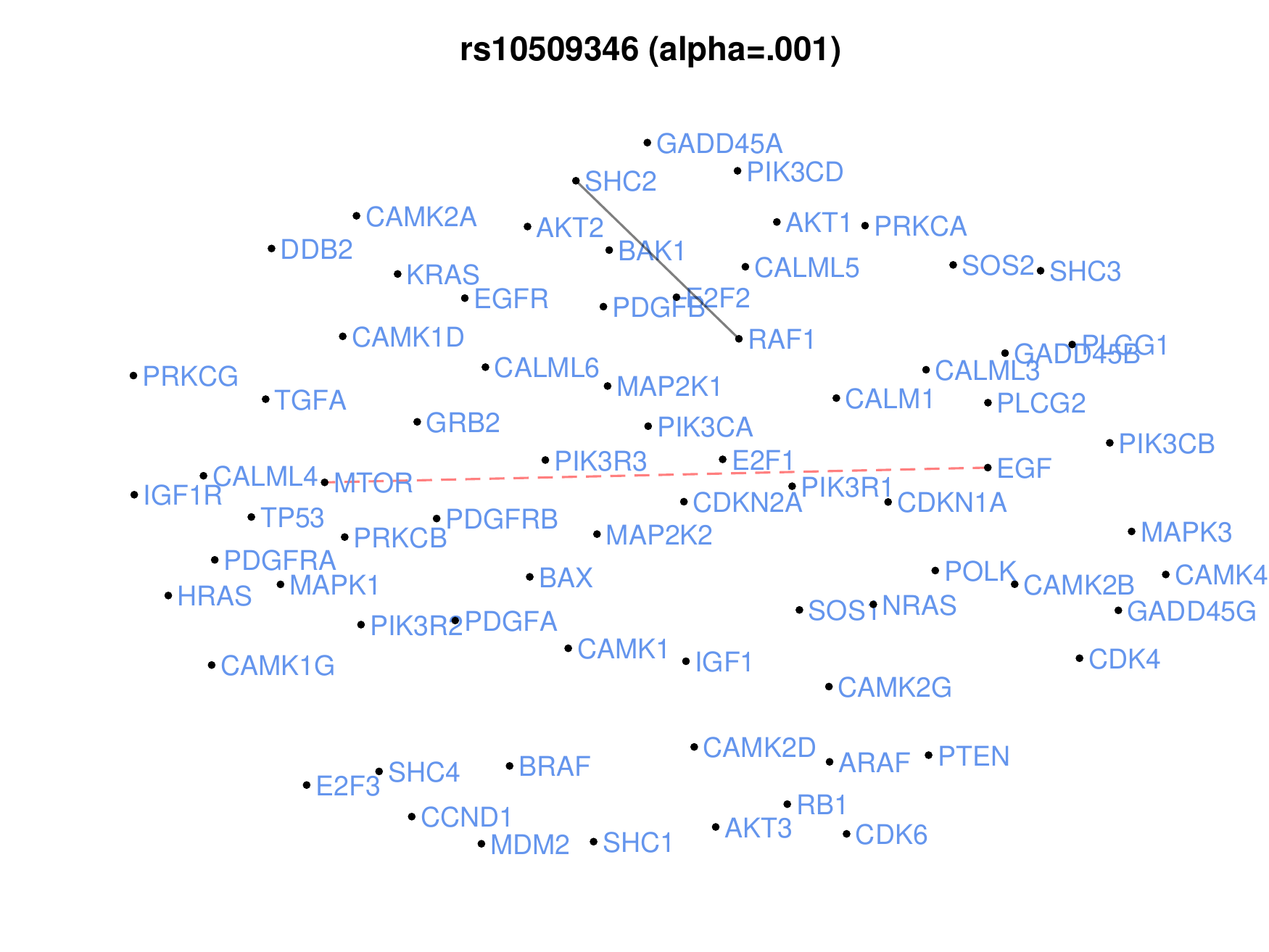}}

\subfigure[Original Selected Edges]{
\label{fig:realh34-origin}
\includegraphics[width=0.31\textwidth]{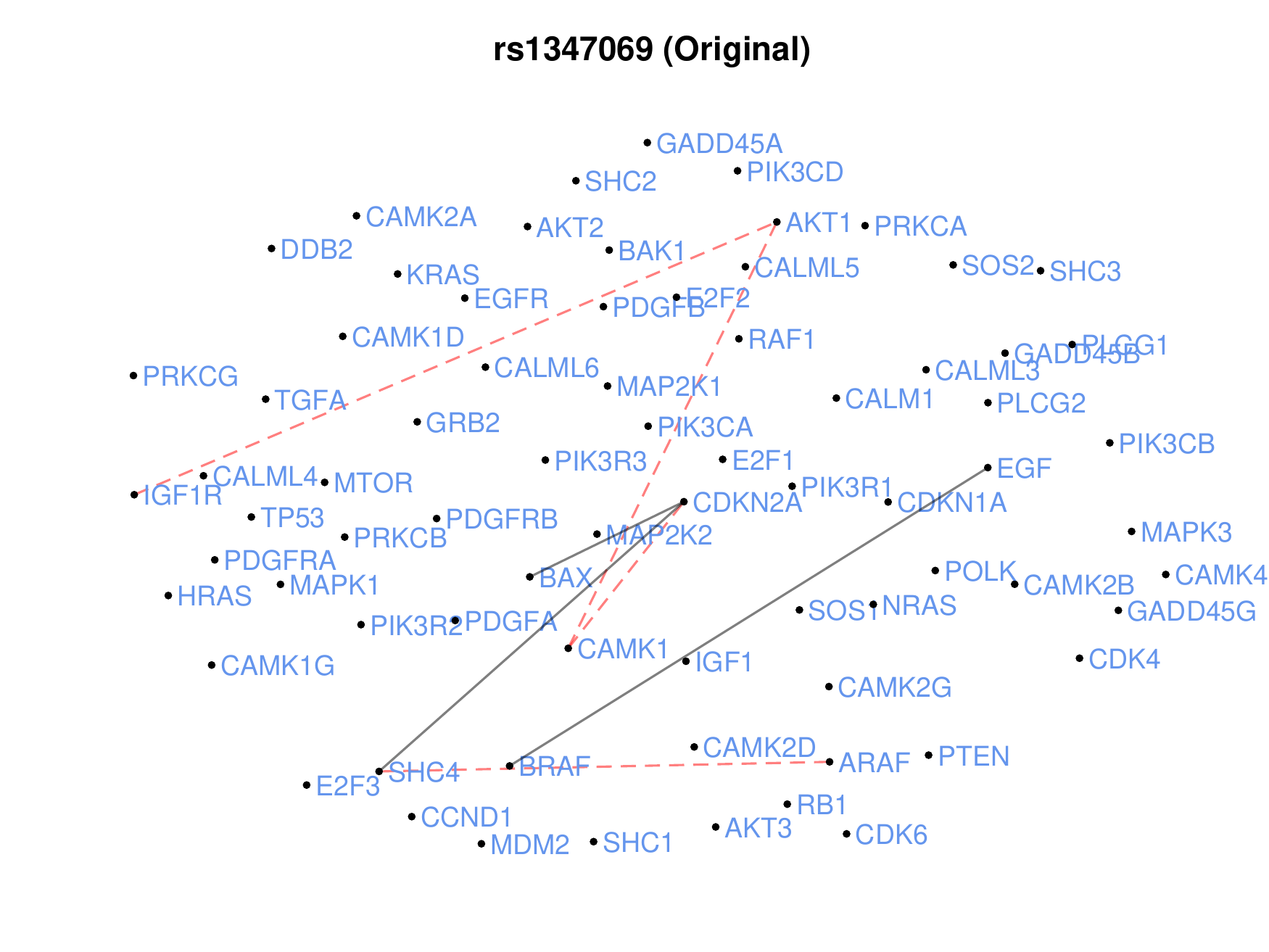}}
\subfigure[Significance Level 0.05]{
\label{fig:realh34-005}
\includegraphics[width=0.31\textwidth]{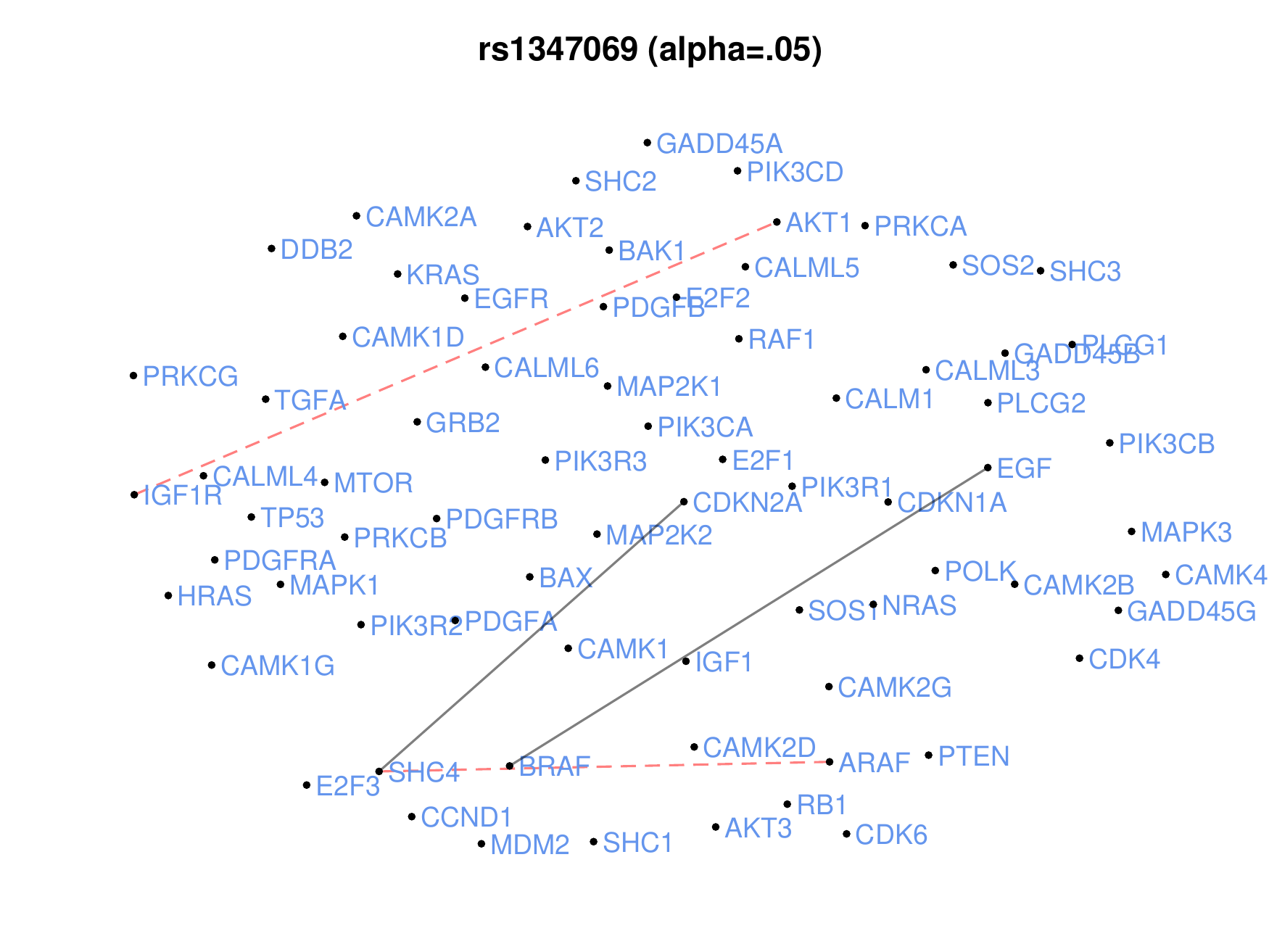}}
\subfigure[Significance Level 0.001]{
\label{fig:realh34-0001}
\includegraphics[width=0.31\textwidth]{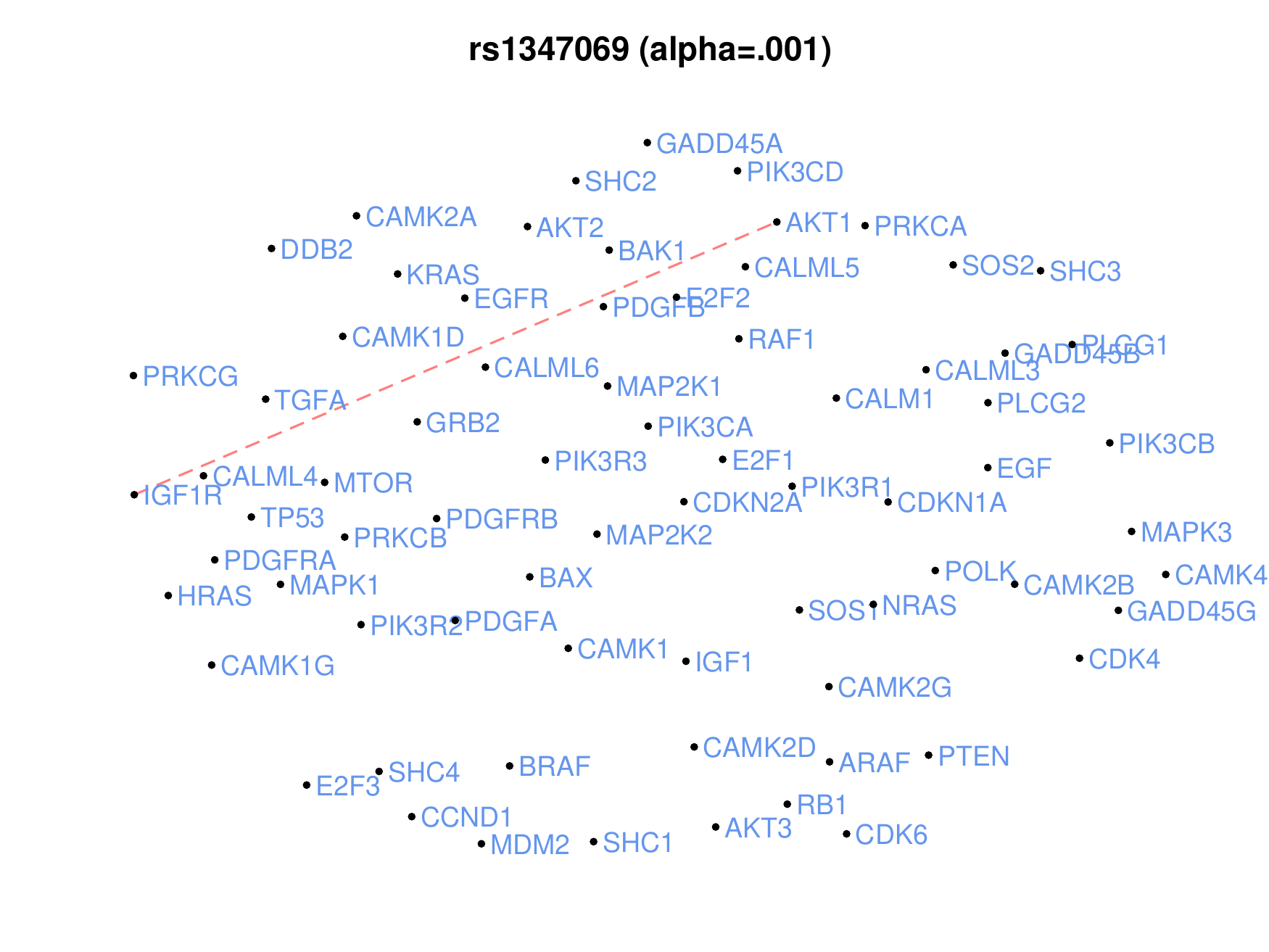}}

\subfigure[Original Selected Edges]{
\label{fig:realh91-origin}
\includegraphics[width=0.31\textwidth]{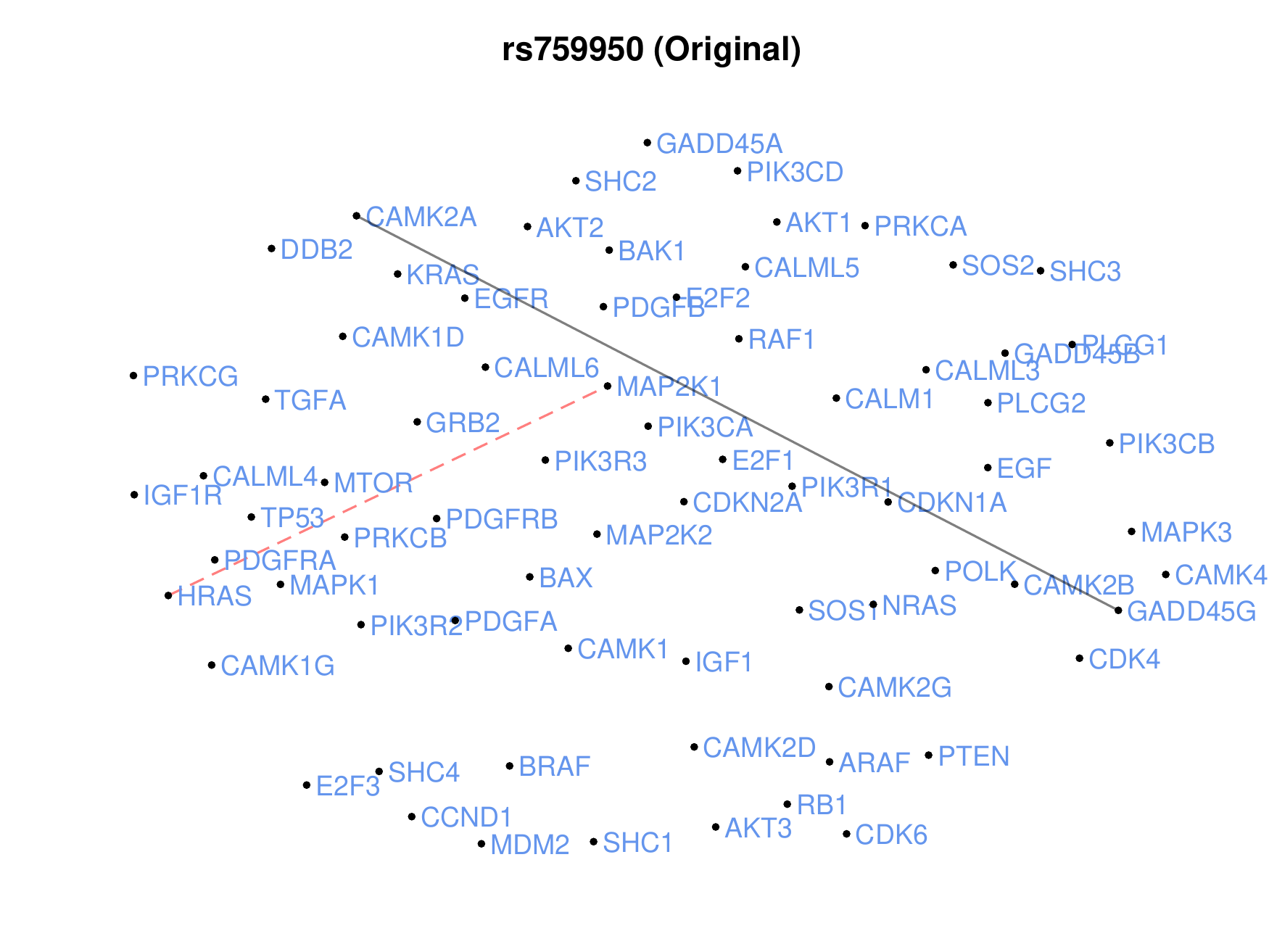}}
\subfigure[Significance Level 0.05]{
\label{fig:realh91-005}
\includegraphics[width=0.31\textwidth]{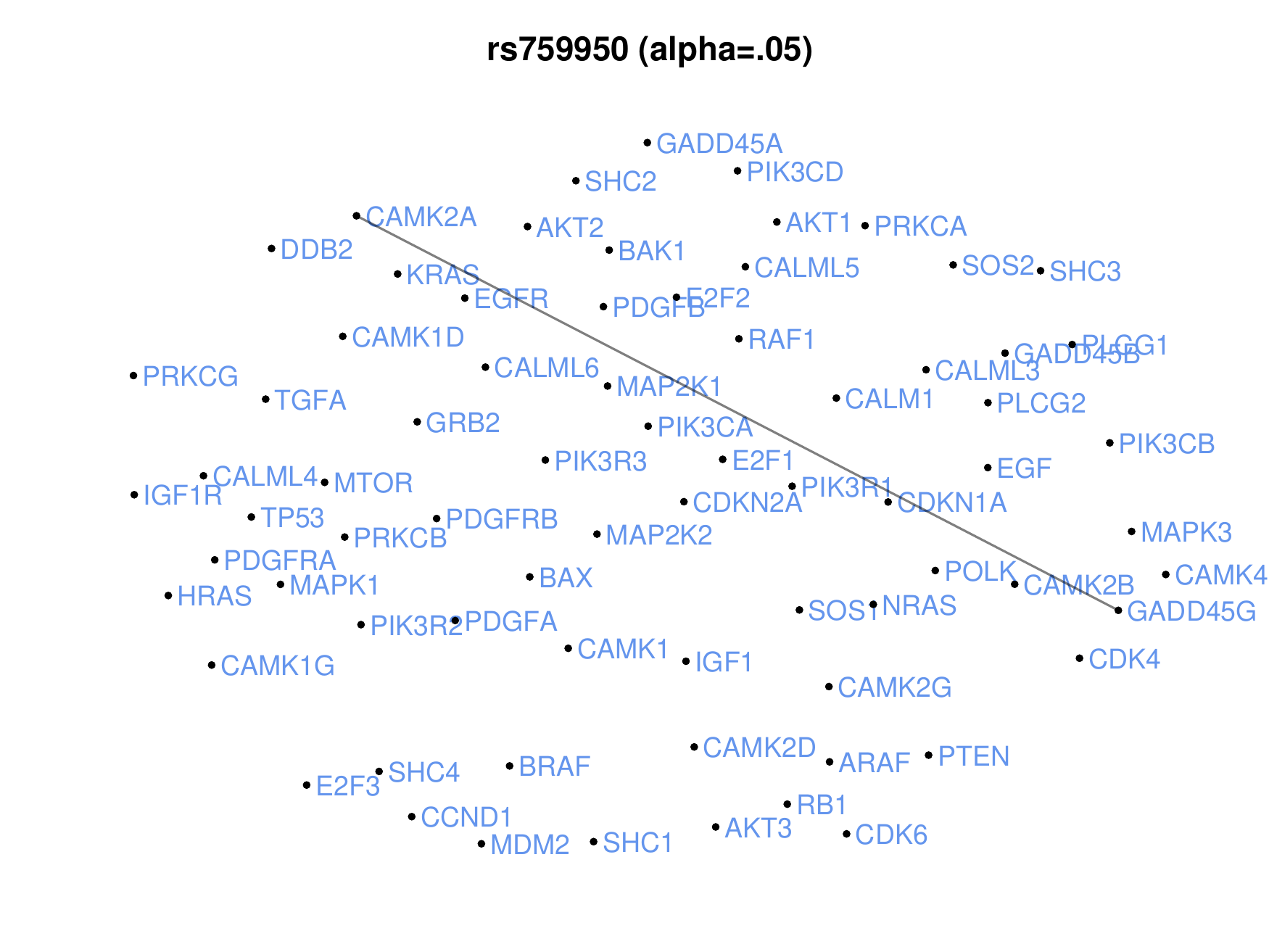}}
\subfigure[Significance Level 0.001]{
\label{fig:realh91-0001}
\includegraphics[width=0.31\textwidth]{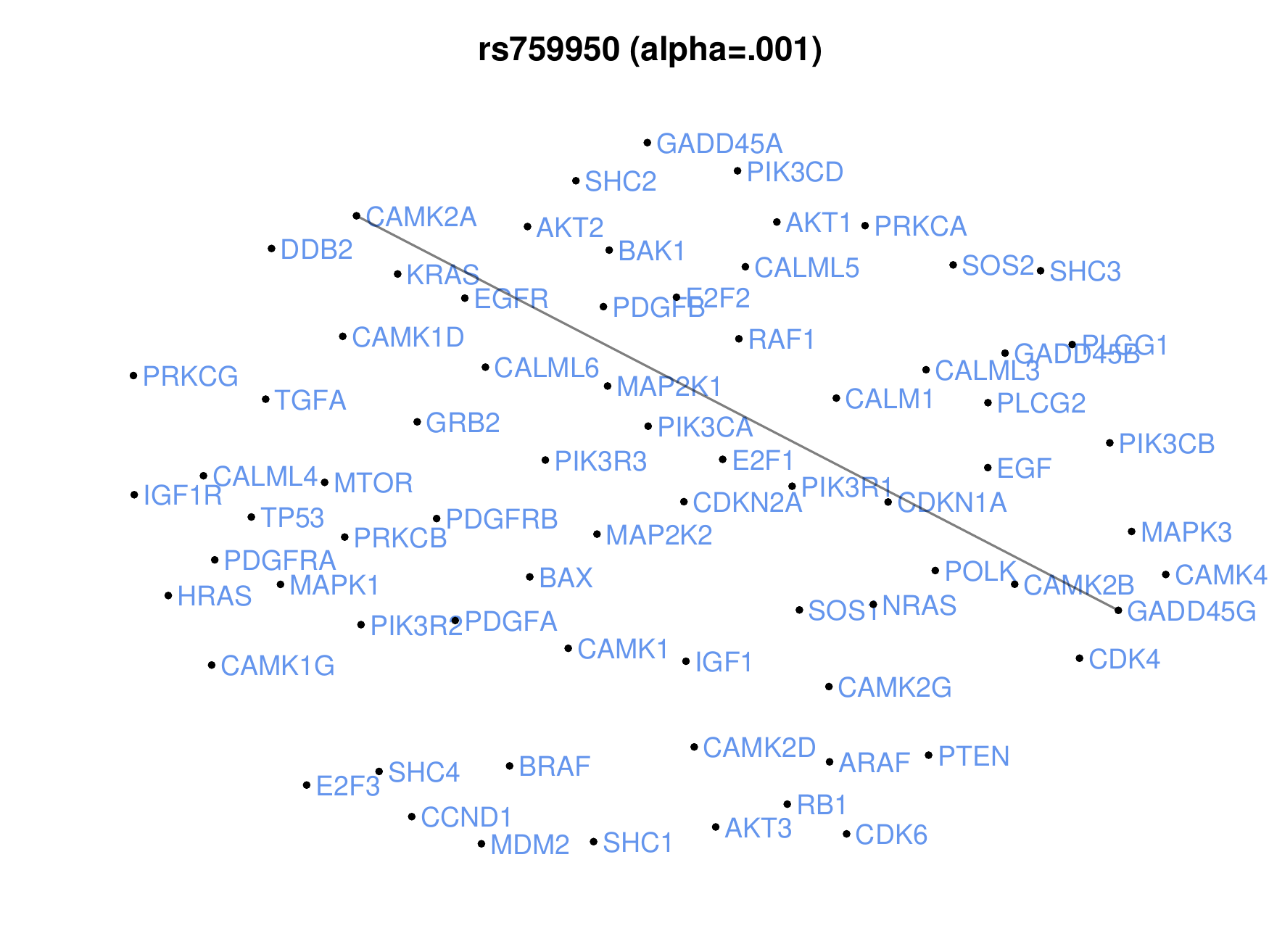}}
{\renewcommand{\baselinestretch}{1.35}\caption{The first, second and third rows display the SNP effects for ``rs10509346'', ``rs1347069'' and ``rs759950'', respectively, at significance levels of 0.05 (middle) and 0.001 (right). Positive partial correlations are shown with red dashed lines, while negative correlations are indicated by black solid lines}
\label{fig:realgraph}}
\end{figure}

Based on the graphical regression results, we further examine the effects of covariates on the graph and identify nine co-expression quantitative trait loci (eQTLs) at a significance level of 0.05: ``rs10509346'', ``rs1347069'', ``rs6701524'', ``rs723210'', ``rs9303511'', ``rs503314'', ``rs728655'', ``rs759950'', and ``rs306098''. At a more stringent significance level of 0.001, three eQTLs remain significant: ``rs10509346'', ``rs1347069'', and ``rs759950''.
 As shown in Figure~\ref{fig:realgraph}, with a significance level of 0.001, we identified a positive correlation between MTOR and EGF and a negative correlation between SHC2 and RAF1 when “rs10509346” is mutated (Figure~\ref{fig:realh19-0001}), both of which are  important in cancer progression \citep{hua2019targeting}. When ``rs1347069'' is mutated (Figure~\ref{fig:realh34-0001}), a positive connection between AKT1 and IGF1R suggests upregulated PI3K/AKT signaling \citep{manning2017akt}. When ``rs759950'' is mutated (Figure~\ref{fig:realh91-0001}),  we identified a negative partial correlation between GADD45G and CAMK2A, indicating potentially opposing roles in cell growth regulation \citep{e2012gadd45,coultrap2012camkii}. We also discover some eQTLs under statistical significance 0.05. For instance, with the mutation of ``rs6701524", there is a negative correlation between PDGFRB and CAMK1, whereas with ``rs503314" mutated there is a positive link between CCND1 and CDKN2A; see Section~\ref{sec:additional_realdata} in the supplement for more results.  Identifying these eQTLs may enable oncologists to tailor treatments to a patient’s unique genetic profile and tumor microenvironment, enhancing efficacy and reducing side effects \citep{li2023identification}.

\section{Conclusions and Discussions}
\label{sec:conclusions}
We  have developed a {\em \underline{s}egmentally \underline{a}djusted  \underline{g}raphical r\underline{e}gression} (SAGE) estimator for multi-task Gaussian graphical regression models, enabling valid statistical inference in high-dimensional settings. In addition,  we have  proposed a  projection approach to  obtain the SAGE estimates,  simplifying the estimation of the inverse variance-covariance matrix and  enhancing computational efficiency. Our theoretical results show that the SAGE estimator asymptotically follows a normal distribution, allowing for reliable confidence intervals and hypothesis testing. The simulation studies and real data analyses confirm the effectiveness of our method, providing valuable insights into biological graphs.

{Our inference method approaches each node separately. For inference on multiple nodes, one could apply the Bonferroni correction or FDR control. However, since these approaches do not directly account for dependence among nodes, they may not achieve optimal power. In contrast, a joint debiasing approach, if feasible, could lead to more efficient inference with reduced variances by leveraging the shared dependence across tasks. However, this presents significant theoretical and computational challenges, which we plan to explore further in the future.}

\bibliographystyle{agsm}
\begingroup
\baselineskip=17.5pt
\bibliography{JASAref}
\endgroup

\end{document}